# Type Checking Extracted Methods

Yuquan Fu[a] and Sam Tobin-Hochstadt[a]

a   Indiana University, Indiana, USA

**Abstract**   Many object-oriented dynamic languages allow programmers to *extract methods* from objects and treat them as functions. This allows for flexible programming patterns, but presents challenges for type systems. In particular, a simple treatment of method extraction would require methods to be contravariant in the receiver type, making overriding all-but-impossible. We present a detailed investigation of this problem, as well as an implemented and evaluated solution.

  Method extraction is a feature of many dynamically-typed and gradually-typed languages, ranging from Python and PHP to Flow and TypeScript. In these languages, the underlying representation of objects as records of procedures can be accessed, and the procedures that implement methods can be reified as functions that can be called independently. In many of these languages, the programmer can then explicitly specify the `this` value to be used when the method implementation is called.

  Unfortunately, as we show, existing gradual type systems such as TypeScript and Flow are unsound in the presence of method extraction. The problem for typing any such system is that the flexibility it allows must be tamed by requiring a connection between the object the method was extracted from, and the function value that is later called.

  In Racket, where a method extraction-like facility, dubbed "structure type properties", is fundamental to classes, generic methods, and other APIs, these same challenges arise, and must be solved to support this feature in Typed Racket. We show how to combine two existing type system features—existential types and occurrence typing—to produce a sound approach to typing method extraction.

  We formalize our design, extending an existing formal model of the Typed Racket type system, and prove that our extension is sound. Our design is also implemented in the released version of Racket, and is compatible with all existing Typed Racket packages, many of which already used a previous version of this feature.

## The Art, Science, and Engineering of Programming

**Type Checking Extracted Methods**

## 1   Methods as Values

> On his next walk with Qc Na, Anton attempted to impress his master by saying "Master, I have diligently studied the matter, and now understand that objects are truly a poor man's closures." Qc Na responded by hitting Anton with his stick, saying "When will you learn? Closures are a poor man's object." At that moment, Anton became enlightened.      —*Anton Van Straaten* [23]

The relationship between objects with methods and functions is fundamental to understanding object-oriented programming and languages. In some languages, such as Smalltalk, "everything is an object", and there is no way to interact with a method except by sending the appropriate message to the object. But in languages ranging from JavaScript [19] to Python [6] to PHP [22] to Racket [12], it is possible to *extract* the function corresponding to a method from its containing object and call it.

Similarly, whether attempting to understand the foundations of object-oriented languages or to implement them on a low-level platform, a standard approach [4] is to encode objects as records of functions, with message sending becoming record selection followed by function call. Thus the necessity of considering methods independently of their containing object arises here too.

The key challenge in all of these settings is the role of **Self**, the receiver of the message. Once a method is separated from its object, there is no longer a designated receiver. This offers new flexibility to programmers, but also the opportunity for error.

In the face of this challenge, languages supporting **method extraction** take two primary approaches. First, systems such as Python avoid the problem entirely by *closing* extracted methods over the object they are extracted from. This ensures that flexibility is not misused by eliminating it entirely. The alternative approach, seen in languages such as JavaScript where object-orientation is built upon records, allows programmers to pick *arbitrary objects* to stand in as the receiver. This flexibility enables new patterns but also makes reasoning about the correctness of programs more challenging and opens up the possibility of hard-to-understand errors.

The challenges of method extraction multiply when combined with type checking, especially in the setting of gradual type systems for existing languages. A naive approach to typing either violates soundness or requires that methods be *both co- and contra-variant* in their receiver type, rendering inheritance impossible. Unfortunately, systems such as TypeScript [15] and Flow [18] have chosen unsoundness, opting to preserve the flexibility apparent without types over a sound approach to the problem.

To show that a synthesis is possible, we present a system which supports method extraction along with sound gradual typing. We work in the context of Typed Racket [11], a mature gradual type system for Racket, an existing untyped language. Racket's *structure type properties* are the focus of our study—they are essentially a vtable-like mechanism for arbitrary records, and building well-typed abstractions from them requires tackling precisely the key challenges posed by method extraction. Structure type properties are also the implementation technology for Racket's system of generic methods, and Racket's Java-like class system.

Our system allows programmers to control when an argument should have the type **Self**, and when it does, then exactly the receiver—that is, the object the method was extracted from—must be supplied in that position. This restriction is the only sound one compatible with both subtyping and inheritance. We show that this restriction is naturally expressed using a combination of existential types [2] and Typed Racket's support for *occurrence typing* [11, 13]. Occurrence typing was originally developed to model predicate tests; here it finds unrelated but compelling use in recording the identity of the receiver.

The resulting system is capable of working with existing idiomatic use of Racket's structure properties, which we validate by implementing the system in Typed Racket and releasing it to all Typed Racket users. Typed Racket had previously and unsoundly *accepted all uses* of structure type properties with no checking whatsoever; with our new implementation virtually all of these unchecked uses simply worked correctly and the remaining cases were both unrelated to our design and easily fixed.

We begin our presentation in section 2 by outlining current approaches to method extraction in gradually-typed languages, starting with JavaScript and its existing type systems. This demonstrates both the problem and the inadequacy of current solutions. We also describe Racket's structure type property system, and how the same problem of method extraction re-occurs in this setting. Section 3 presents our type-checking approach at a high level, focusing on examples. In section 4, we present a formal model of our approach, extending existing models of Typed Racket, which we prove sound. Section 5 describes our implementation, including how we dynamically enforce the new types we have added, as required for gradual typing. We then discuss the practical experience gained by applying our system, now released in the current version of Racket, to existing Typed Racket programs. Finally, we compare with related approaches and conclude.

## 2 The Current State of the Art

Numerous dynamic languages both support method extraction and have recently developed gradual type systems, among them JavaScript, Python, PHP, and Ruby[26]. With the exception of Python, discussed below, all of them fail to soundly enforce the type system in the presence of extracted methods.

Since JavaScript is the language with both the most mature type systems and the simplest model of objects as records with functions as members, we first present the issues in that context. Then we show how the same issues reoccur in Racket's structure type properties.

### 2.1 Unsound Method Extraction in JavaScript

We begin with the key issue—the type of `self` or `this` for extracted methods in the presence of inheritance. Consider the program in listing 1, which uses the syntax accepted by both TypeScript and Flow for type annotations.

In this program, we have two classes, one for two-dimensional coordinates and an extension for three dimensions. Each defines a simple constructor that initializes its

**Type Checking Extracted Methods**

fields and a `dist` method that computes the distance between the current point and a specific point.

We then construct an instance of each, but annotate p3d, the three-dimensional instance, with the `Point2D` type. We also extract the `dist` method from p3d using JavaScript's `.bind()` method with the `this` keyword mistakenly set to the two-dimensional instance p2d. Then we intend to use the extracted method to calculate distance between p3d and every element of an array of three-dimensional points.

■ **Listing 1** Unsound Method Extraction in TypeScript and Flow

```
class Point2D {
  x : number; y : number;
  constructor(x : number, y : number) {
    this.x = x; this.y = y;
  }
  dist(this:this, target: Point2D) {
    return Math.sqrt(Math.pow(target.x - this.x, 2) +
          Math.pow(target.y - this.y, 2));
  }
}

class Point3D extends Point2D {
  z : number;
  constructor(x : number, y : number, z : number) {
    super(x, y); this.z = z;
  }

  dist(this:this, target : Point2D) : number {
    if (target instanceof Point3D) {
      return Math.sqrt(Math.pow(target.x - this.x, 2) +
        Math.pow(target.y - this.y, 2) +
        Math.pow(target.z - this.z, 2));
    } else {
      throw "Target is not a Point3D";
    }
  }
}

var p2d = new Point2D(0, 0);
var p3d : Point2D = new Point3D(3, 4, 5);
var meth = p3d.dist.bind(p2d);
[new Point3D(5, 6, 7), new Point3D(9, 10, 11)].map(meth);
```

We would hope that the type checkers would reject the program, since `meth` is the extracted method `dist` of a `Point3D` instance and only works if the receiver is an instance of `Point3D` or its subtype. However, the type checkers report no errors. When we run the program, the p2d is used as `this` in the method `Point3D`'s `dist` and hence `this.z` evaluates to `undefined`. Because of JavaScript's type coercion for arithmetic operations, this produces `NaN` rather than an error, but is straightforwardly unsound.

The fundamental issues for TypeScript and Flow are different. In Flow, `.bind` can take in *any* value as the first argument, regardless of the connection to `Point3D`. This preserves maximum flexibility, but will not do for a sound type system. On the other hand, TypeScript's `.bind` accepts a value of any *subtype* of the current class for `this`, and the type of `p3d` when `dist` is extracted is `Point2D`.

More generally, it is well known that method overriding is sound when arguments vary *contravariantly* in the subclass. Of course, the type of `this` varies *covariantly* in subclasses—that's what it means for subclasses to be subtypes. But method extraction makes the `this` parameter into an argument, producing a contradiction that leads to unsoundness if not addressed.

## 2.2 Structures and Structure Type Properties

Having seen the challenge of method extraction in JavaScript, we now turn to our setting—Racket structures and structure type properties, and how they face all of the challenges of method extraction in a way that requires a full solution. We begin with an overview, and then rediscover the same problems.

Racket's *structures* are records with named fields and inheritance defined with the `struct` form:

◼ **Listing 2** Two and Three-Dimensional Points in Racket

```
1  (struct point2d [x y]
2    #:property prop:how-big 10
3    #:property prop:custom-write
4    (lambda (self)
5      (printf "Point(~a, ~a)" (point2d-x self) (point2d-y self))))
6
7  (struct point3d point2d [z]
8    #:property prop:custom-write
9    (lambda (self)
10     (printf "Point(~a, ~a, ~a)" (point2d-x self) (point2d-y self)
11             (point3d-z self))))
```

As shown in the code above, a `struct` form also introduces several names to the current scope. For example, the `struct` form for `point2d` at least defines the following names: 1. `point2d`, a constructor procedure to create an instance of `point2d` with a value for each field defined in it; 2. `point2d?`, a predicate procedure to check if an arbitrary value is a `point2d`, producing a boolean; 3. `point2d-x` and `point2d-y`, two field accessor procedures that take a `point2d` and produce the values of the field x and y respectively.

Structure definitions can also inherit from other structure types; doing so means that new fields are additive and that instances of the structure subtype are treated as instances of the supertype. Note that structure subtyping in Racket is nominal. In the code above, we create the structure `point3d` based on the structure `point2d`.

Structures also support *structure type properties*, a per-type map of property keys to arbitrary values. In listing 2, the `point2d` structure has two structure type properties: `prop:how-big`, whose value is 2, and `prop:custom-write`. The value supplied for

**Type Checking Extracted Methods**

`prop:custom-write` is a function whose first argument is expected to be an instance of the structure type.[1]

As with structures, a structure type property is defined by a collection of generated functions and values specific to that property. These values are created by the function `make-struct-type-property`, which takes a symbol naming the property and returns three values: a property descriptor to be used in a `struct` definition, as well as a predicate procedure and an accessor procedure:

```
1 (define-values (prop:custom-write custom-write? custom-write-accessor)
2    (make-struct-type-property 'custom-write))
```

In the code above, the predicate procedure `custom-write?` returns true if the argument is an instance of any structure type with a value attached for the corresponding property `prop:custom-write`. The accessor procedure `custom-write-accessor` extracts the property value paired with the property descriptor of a structure type from its instance, raising an error for values that don't have the relevant property.

The design of structure type properties makes them similar to Java static fields: there is a single value per-type but that value is accessible from individual instances. The key distinction is that access to structure type properties is mediated by values that serve as capabilities: the accessor and property descriptor.

Structure type properties allow defining extensible abstractions, such as the following customizable printer `print-value`:

```
1 (define (print-value v)
2   (if (custom-write? v)
3       ((custom-write-accessor v) v)
4       (printf "unknown value")))
```

First, it checks if the input `v` has a value for the `custom-write` property, using the corresponding predicate. If so, that property value is extracted from `v` with the appropriate accessor and used to print the value, by passing `v` to the custom printing function.

This example demonstrates one common pattern used with structure type properties: the property serves as a single-entry vtable, and an abstraction around the property defines a generic function which supplies the appropriate `self` value.

Modular encapsulation allows the use of `custom-write-accessor` to be limited to just the `print-value` function, ensuring that the value supplied in the `struct` definition is not misused, perhaps by passing some other value as the input. However, Racket programmers can easily break the invariant by making a similar mistake to what we have shown in listing 1:

```
1 (define (print-value2 v1 v2)
2   (if (and (custom-write? v1) (custom-write? v2))
3       ((custom-write-accessor v1) v2)
4       (printf "unknown value")))
```

---

[1] The actual `custom-write` property in Racket is somewhat more complex, in ways that are not relevant to our discussion.

In `print-value2`, we extract the print method from `v1`, but then invoke it with `v2`. Suppose we supply `v1` with a `point3d` value and `v2` with a `point2d` value. Though both `v1` and `v2` pass the check, the function will raise a runtime error, because the `point2d` value cannot be applied to the method `point3d-z` used in the structure `point3d`'s property value for `prop:custom-write`.

### 2.3 A View to Solutions

Other dynamic languages with gradual type systems face versions of this problem, and have addressed it in different ways. Hack [21], a typed version of PHP, has evolved into a new language and dispensed with method extraction. Sorbet [24], for Ruby, seems to have a similar approach to Flow.

However, two other systems take a different tack. In Java [8], when using reflection [20] to extract a method, the resulting value tracks the runtime class it was extracted from, and requires an instance of that class to be supplied at invocation time. Python, when extracting a method, *closes* the method over the object it is extracted from, instead of allowing it to be supplied later.

The key difference between Java and Python, on the one hand, and systems such as Racket and JavaScript, where the problem arises, is access to the underlying view of objects as records of functions. In Racket, where it is implemented via macros, and in JavaScript, where it is built into the semantics of method call syntax, method calls are simply *patterns of use* of this lower-level view. While it is possible for JavaScript to adopt the treatment of extracted methods in Python at the cost of backward-incompatibility, it is impossible for Racket to redesign structure type properites in a similar fasion, because they have a variety of use cases and serving as a method table is just one of them.

In such languages, type systems should support the mechanisms directly, and not merely certain patterns of use that fit in predefined categories, such as type checking of Racket's class system [9, 14]. As we will see, an appropriate static type system can preserve soundness while keeping the original runtime behavior.

## 3 Types for Structure Type Properties

With an understanding of structures and properties in Racket in hand, we now describe our approach for typing these features, including the key issue of what the legal arguments to the function attached to the `custom-write` property are.

### 3.1 Declaring Typed Structure Properties

**Listing 3** Typed Structure Point

```
1  (struct point2d ([x : Integer] [y : Integer])
2    #:property prop:how-big 10
3    #:property prop:custom-write
4    (lambda ([self : point2d]) : Void
5      (printf "Point(~a, ~a)" (point2d-x self) (point2d-y self))))
```

**Type Checking Extracted Methods**

Consider a typed version of the structure definition shown previously, in Typed Racket syntax, given in listing 3. This definition follows the similar one in untyped Racket closely, with type annotations in three places: on the two fields, `x` and `y`, and on the argument to the custom printer, which takes a `point2d`, as expected.

In order for this to work, the structure type property descriptors must be equipped with types; for example `(Struct-Property Number)` for `prop:how-big`, using a new unary type constructor `Struct-Property`. But for structure type properties like `prop:custom-write`, the type is less obvious. Obviously it cannot already have the type `(Struct-Property (-> point2d Void))`, since the function type `(-> point2d Void)` only works for the initial definition of `point2d`. And yet, in the body of the value expression, `self` must be of type `point2d`, since it must be a suitable input to `point2d-x`.

Our solution is a new built-in type **Self**, denoting the type of the structure that the property declaration is embedded in, `point2d`. Thus the type of `prop:custom-write` can be expressed as `(Struct-Property (-> Self Void))`.

To type check the declaration in listing 3, the type checker simply substitutes the actual structure type, `point2d`, for **Self**.

## 3.2 Type Refinement with Predicates

The next challenge is the definition of `print-value`. First, what should the domain of the `custom-write-accessor` function be? It must be restricted in some way, yet open to further extension. We represent this with a new type constructor, dubbed `Has-Struct-Property`, which allows the domain to be `(Has-Struct-Property prop:custom-write)`.

The next challenge is that `print-value` is intended to work on *all* inputs, not just those with the property set—that's why it has a predicate test at all. Fortunately, Typed Racket comes with a pre-existing solution to this problem: occurrence typing. This is an approach that enables the type system to obtain type information about its argument from a predicate procedure and then propagate that information to branches of control flow. For example, in `(if (number? v) v 17)`, `v` initially might have the type Any, just as the parameter to `print-value` does. Occurrence typing refines `v` to the type `Number` in then branch, which is the logical corollary of `(number? v)` evaluating to `t`. This is expressed by giving the `number?` predicate the type `(-> Any Boolean : Number)`. The last part is a *latent proposition* that the input must be a number if the function produces a true value.

Similarly, the type `(-> Any Boolean : (Has-Struct-Property prop:custom-write))` for `custom-write?` allows the `print-value` function to type check.

## 3.3 Structure Type Property Access Is Method Extraction

We now turn to the *result* of `(custom-write-accessor v)`, which is precisely an extracted method. Based on the type of the function associated with the property in listing 3, the result should have type `(-> Self Void)`. Furthermore, to make the outer application type check, `v` must then have type **Self**.

Let us consider a few possible options. One obvious choice is to replace **Self** with (`Has-Struct-Property prop:custom-write`). Then the entirety of `print-value` will type check. Unfortunately, this is not sound—it's the same problem that we saw for TypeScript. Any other value with the correct property would then be allowed instead of v, even one that was not an instance of `point2d`.

Instead, we turn to *existential types*. We existentially quantify over **Self**, allowing only values that we know to be appropriate as an argument to the extracted method.

However, existential types are not enough by themselves. If we gave the extracted method the type (`Some (Self) (-> Self Void)`), our system would be sound, but our method would be impossible to apply!

One possible strategy is to change the semantics of structure type property access. An accessor could produce a package of both the receiver value and the extracted property value, with an existential type connecting the two. Then the type of `custom-write-accessor` is (`-> (Has-Struct-Property prop:custom-write) (Some (X) (Pairof X (-> X Void)))`) and adding an unpacking operation to the language for existential types: (`let-unpack ([(X x) e]) b`) where e has type (`Some (X) S`). For method extraction, x has a pair of the unpacked instance and an extracted function. The function call in `print-value` would become:

```
(let-unpack ([(X (new-v meth)) (custom-write-accessor v)])
            (meth new-v))
```

However, this would require invasive changes to Racket's runtime system as well as backwards-incompatible changes to all existing uses of structure type properties—the opposite of the goals of gradual type systems such as Typed Racket.

### 3.4 Combining Existential Types and Occurrence Typing

To solve this problem, we extend the existential type approach in two ways. First, we automatically and implicitly unpack the existential at the point where the `custom-write-accessor` function is applied. Second, we use Typed Racket's support for type refinement to refine the type of v to be **Self**.

We have already seen type refinement for `number?` or `custom-write?`, but here instead of refining based on a predicate, we refine the type of the argument to the accessor function to have the type of the existentially-quantified variable.

The resulting type is (`-> (Has-Struct-Property prop:custom-write) (Some (X) (-> X Void) : X)`). Here X appears not just in the domain of the method but also in the proposition, stating that after we've applied the function, we know that the input has type X.

By putting all the types above together, we can give types to the generated structure type property descriptor, predicate procedure, and accessor procedure when creating a new structure type property:

```
(: prop:custom-write (Struct-Property (-> Self Void))
(: custom-write?
   (-> Any Boolean : (Has-Struct-Property prop:custom-write)))
(: custom-write-accessor
   (-> (Has-Struct-Property prop:custom-write) (Some (X) (-> X Void) : X)))
(define-values (prop:custom-write custom-write? custom-write-accessor)
               (make-struct-type-property custom-write))
```

Furthermore, these types now allow us to type check the `print-value` function, exactly as originally written.

## 4 Formal Model

Our calculus $\lambda_{ETR}$ extends $\lambda_{TR}$ [13], a formal model of Typed Racket. The presentation of the formal model starts with the introduction of the typing judgment, followed by the descriptions of the syntax and the novel typing rules. In section 4.2.1, we discuss the generativity of **let-struct** and **let-struct-property**. Section 4.2.2 describes an illustrative example to demonstrate how the new typing rules work. We then present a soundness proof for our calculus in section 4.3.

The fundamental judgment of our type system is:

$$\Gamma \vdash e : (\tau \,;\, \psi_+ \,|\, \psi_- \,;\, o)$$

It states that in the type environment $\Gamma$ four properties of the expression $e$ hold:

- $e$ has the type $\tau$
- if $e$ evaluates to a non-false value, the *true proposition* $\psi_+$ holds
- otherwise, the *false proposition* $\psi_-$ holds.
- $o$ is a *symbolic object* referencing a portion of a *runtime* environment. If $o$ is not $\emptyset$ (the null object), looking it up in the runtime environment produces the same value as evaluating $e$.

### 4.1 Syntax

The syntax of terms, values, types, propositions, objects, and environments are given in figure 1, where new forms are highlighted.

**Expressions** Our system supports conditionals, let-binding, numeric and boolean constants, abstraction with a typed parameter, application, pairs and field accesses to them as well as primitive operations. **let-struct** creates a structure with specified name $sn$, a field type $\tau$, a collection of structure type property names and their value expressions $\overrightarrow{sp\,e}$, and it binds three identifiers to a structure constructor procedure, a structure predicate procedure and a structure field accessor procedure for use in the body $e$. **let-struct-property** creates a structure type property, which is named $x$ and has a value type $\tau$. It also introduces the following three identifiers among others to the body $e$: a property descriptor, a predicate and an accessor procedure for the property. In our system, we chose to include the names of structures as parts of their types. This design decision reflects the fact that structure types in Racket are nominal, and thus two structures with different names are not considered identical when their fields and properties are equal. Note that **let-struct** is generative, while **let-struct-property** is not. See a detailed discussion in section 4.2.1.

**Values** Besides the values seen in $\lambda_{TR}$, we also added structure instances, structure type property descriptors and their companion procedures. $sn(v : \tau, \overrightarrow{sp\,v})$ describes

$$
\begin{array}{ll}
op ::= \text{not} \mid \text{add1} \mid \text{nat?} \mid ... & \textbf{Primitive Ops} \\
e ::= & \textbf{Expressions} \\
\quad \mid x \mid sn \mid sp & \text{variable} \\
\quad \mid n \mid \text{true} \mid \text{false} \mid op & \text{base values} \\
\quad \mid \lambda x{:}\tau.e \mid (e\ e) & \text{abstraction, application} \\
\quad \mid (\textbf{if}\ e\ e\ e) & \text{conditional} \\
\quad \mid (\textbf{let}\ (x\ e)\ e) & \text{local binding} \\
\quad \mid (\textbf{let-struct}\ ((x\ x\ x)\ (sn\ \tau\ \overrightarrow{(sp\ e)}))\ e) & \text{structure binding} \\
\quad \mid (\textbf{let-struct-property}\ ((sp\ x\ x)\ (x\ \tau)))\ e) & \text{structure property binding} \\
\quad \mid (\textbf{cons}\ e\ e) & \text{pair construction} \\
\quad \mid (\textbf{fst}\ e) \mid (\textbf{snd}\ e) & \text{field access} \\
v ::= & \textbf{Values} \\
\quad \mid n \mid \text{true} \mid \text{false} \mid op & \text{base values} \\
\quad \mid \langle v,v \rangle \mid [\rho, \lambda x{:}\tau.e] & \text{pair, closure} \\
\quad \mid sn(v : \tau, \overrightarrow{sp\ \vec{v}}) & \text{structure instance} \\
\quad \mid \text{pd}(sp) & \text{struct property descriptor} \\
\quad \mid so & \text{struct related operations} \\
so := & \textbf{Struct-Related-Operations} \\
\quad \mid \text{ctor}(x, \tau, \overrightarrow{sp\ \vec{v}}) \mid \text{pred}(sn(\tau, \overrightarrow{sp})) \mid \text{acc}(sn(\tau, \overrightarrow{sp})) & \text{ops for structure instance} \\
\quad \mid \text{p-pred}(sp) \mid \text{p-acc}(sp, \tau) & \text{ops for structure properties} \\
\tau, \sigma ::= & \textbf{Types} \\
\quad \mid \top & \text{universal type} \\
\quad \mid \mathbf{N} \mid \mathbf{T} \mid \mathbf{F} \mid \tau \times \tau & \text{basic types} \\
\quad \mid (\bigcup \vec{\tau}) & \text{untagged union type} \\
\quad \mid sn(\tau, \overrightarrow{sp}) & \text{struct type} \\
\quad \mid \textbf{Prop}(\tau) & \text{struct property type} \\
\quad \mid \textbf{Has-Prop}(sp) & \text{has struct property type} \\
\quad \mid \textbf{Self} & \text{the receiver type} \\
\quad \mid X & \text{type variable} \\
\quad \mid \exists X.\, x{:}\tau \to R & \text{existential function type} \\
\psi ::= & \textbf{Propositions} \\
\quad \mid \mathbb{TT} \mid \mathbb{FF} & \text{trivial/absurd prop} \\
\quad \mid o \in \tau \mid o \notin \tau & \text{atomic prop} \\
\quad \mid \psi \wedge \psi \mid \psi \vee \psi & \text{compound props} \\
\varphi ::= \text{fst} \mid \text{snd} & \textbf{Fields} \\
o ::= & \textbf{Symbolic Objects} \\
\quad \mid \emptyset & \text{null object} \\
\quad \mid x & \text{variable reference} \\
\quad \mid (\vec{\varphi}\ o) & \text{object field reference} \\
\xi ::= \psi \mid sp & \textbf{Environment Elements} \\
R ::= (\tau\ ;\ \psi \mid \psi\ ;\ o) & \textbf{Type Result} \\
\Gamma ::= \vec{\xi} & \textbf{Environments} \\
\rho ::= \overrightarrow{x \mapsto \vec{v}} & \textbf{Runtime Environments}
\end{array}
$$

**Figure 1** $\lambda_{ETR}$ Syntax

**Type Checking Extracted Methods**

that an instance is created from a structure named *sn* with the field value *v* of type $\tau$, and the instance also inherits a collection of property names and property values from the structure. The constructor procedure ctor(*sn*, $\tau$, $\overrightarrow{sp\,v}$) is used to create an instance for a structure named *sn* with the field of type $\tau$ and a collection of property names and property values. The predicate procedure pred(*sn*) checks if a value is an instance of the structure *sn*. The field accessor procedure acc($sn(\tau, \overrightarrow{sp})$) obtains the field value from an instance of the structure *sn*. The property predicate procedure p-pred(*sp*) checks if a value is an instance of a structure with the property *sp*. The property accessor procedure p-acc(*sp*, $\tau$) takes an instance of a structure with the property *sp*, and returns the value associated with the structure for the property *sp*.

**Types**  The system supports the supertype of all types, the $\top$ type. **N** is the type of all numeric expressions. T and F are the types of all expressions that evaluate to true and false respectively. $\tau \times \tau$ describes a pair type. The untagged union type ($\bigcup \vec{\tau}$) is a supertype of its component. For convenience, the boolean type **B** is the abbreviation of ($\bigcup$ T F). To simplify our exposition, structures in our system have only one field. Structure types are written $sn(\tau, \overrightarrow{sp})$, where *sn* is the name, $\tau$ is the field's type and $\overrightarrow{sp}$ represents a collection of structure property names. **Prop**($\tau$) is the type of a structure type property descriptor, and $\tau$ specifies the type of the expected property values supplied in a structure's definition. Type **Has-Prop**(*sp*) stands for a collection of structure types attached with the property *sp*. **Self**, only used in **Prop**($\tau$), denotes the receiver type. When the quantifier $X$ isn't referenced in the body of the existential function type $\exists X. x{:}\tau \to R$, we abbreviate it to $x{:}\tau \to R$. In our system, an existentially functional value doesn't require explicitly unpacking.

**Propositions**  Propositions, borrowed from propositional logic, are key components of our system. $\mathbb{TT}$ is the trivial proposition and $\mathbb{FF}$ the absurd proposition. The atomic proposition states whether a symbolic object has the type $\tau$. The two operations for compound propositions $\wedge$ and $\vee$ are for conjunction and disjunction of propositions respectively.

Our system uses *fields* to access pair-encoding structural values. *Symbolic objects* denotes portion of runtime-environment.

The *type environments* are extensions to standard type environments. In addition to variables' type information, they also include *propositions* and created structure property names.

The *runtime environments* are standard mappings between variables and their closed runtime values.

## 4.2 Typing Rules

Since most typing rules in $\lambda_{ETR}$ are the same as in $\lambda_{TR}$, we will only show extensions. See appendix A for the full definition.

**Structure Related Values**  T-Property-Descriptor shows the named property pd(*sp*) has type **Prop**($\tau$). T-Struct-Instance shows the structure instance $sn(v : \tau, \overrightarrow{sp\,v_p})$

T-Property-Descriptor

$\Gamma \vdash \mathsf{pd}(sp) : (\mathbf{Prop}(\tau) ; \mathbb{TT} \mid \mathbb{FF} ; \emptyset)$

T-Struct-Related-Operations

$\Gamma \vdash so : (\Delta_s(so) ; \mathbb{TT} \mid \mathbb{FF} ; \emptyset)$

T-Struct-Instance

$\Gamma \vdash sn(v : \tau, \overrightarrow{sp\ v_p}) : (sn(\tau, \overrightarrow{sp}) ; \mathbb{TT} \mid \mathbb{FF} ; \emptyset)$

T-Let-Struct-Property

$$\frac{\begin{array}{c} \tau_p = \mathbf{Prop}(\tau) \\ \tau_{pred} = x{:}\top \to (\mathbf{B} ; x \in \mathbf{Has\text{-}Prop}(sp) \mid x \notin \mathbf{Has\text{-}Prop}(sp) ; \emptyset) \\ \tau_a = x{:}\mathbf{Has\text{-}Prop}(sp) \to \exists X. (\tau[\mathbf{Self} \Mapsto X] ; x \in X \mid \mathbb{TT} ; o_3) \\ \Gamma, sp, x_p \in \tau_p, x_{pred} \in \tau_{pred}, x_{acc} \in \tau_a \vdash e : R \\ X \# \Gamma\ \ X \# sp \# \Gamma \end{array}}{\Gamma \vdash (\mathbf{let\text{-}struct\text{-}property}\ ((x_p\ x_{pred}\ x_{acc})\ (sp\ \tau)))\ e) : R[sp \Mapsto \emptyset][x_{pred} \Mapsto \emptyset][x_{acc} \Mapsto \emptyset]}$$

T-Let-Struct

$$\frac{\begin{array}{c} \overrightarrow{\Gamma \vdash sp : (\mathbf{Prop}(\tau_p) ; \mathbb{TT} \mid \mathbb{FF} ; \emptyset)} \\ \overrightarrow{\Gamma \vdash e_p : (\tau_p[\mathbf{Self} \Mapsto sn(\tau, \overrightarrow{sp})] ; \psi_+ \mid \psi_- ; o_1)} \\ \tau_c = x{:}\tau \to (sn(\tau, \overrightarrow{sp}) ; \mathbb{TT} \mid \mathbb{FF} ; \emptyset) \qquad \tau_a = x{:}sn(\tau, \overrightarrow{sp}) \to (\tau ; \mathbb{TT} \mid \mathbb{FF} ; \emptyset) \\ \tau_p = x{:}\top \to (\mathbf{B} ; x \in sn(\tau, \overrightarrow{sp}) \mid x \notin sn(\tau, \overrightarrow{sp}) ; \emptyset) \\ \Gamma, x_{ctor} \in \tau_c, x_{pred} \in \tau_p, x_{acc} \in \tau_a \vdash e : R \end{array}}{\Gamma \vdash (\mathbf{let\text{-}struct}\ ((x_{ctor}\ x_{pred}\ x_{acc})\ (sn\ \tau\ \overrightarrow{(sp\ e_p)}))\ e) : R[x_{ctor} \Mapsto \emptyset][x_{pred} \Mapsto \emptyset][x_{acc} \Mapsto \emptyset]}$$

T-Abs

$$\frac{\Gamma, x \in \tau \vdash e : R}{\Gamma \vdash \lambda x{:}\tau.e : (\exists X. x{:}\tau \to R ; \mathbb{TT} \mid \mathbb{FF} ; \emptyset)}$$

T-App

$$\frac{\begin{array}{c} \Gamma \vdash e_1 : (x{:}\tau \to \exists X.R ; \psi_{1+} \mid \psi_{1-} ; \emptyset) \\ \Gamma, \psi_{1+} \vdash e_2 : (\sigma ; \psi_{2+} \mid \psi_{2-} ; o_2) \\ \Gamma \vdash \sigma <: \tau \quad X \# \sigma \quad X \# \psi_{1+} \quad X \# \Gamma \end{array}}{\Gamma \vdash (e_1\ e_2) : R[x \stackrel{\sigma}{\Mapsto} o_2]}$$

**Figure 2** Extension of Typing Rules

has type $sn(\tau, \overrightarrow{sp})$. Through the metafunction $\Delta_s$, T-Struct-Related-Operations assigns function types to primitive operations for structure instances and structure type properties. Figure 3 describes the definition of $\Delta_s$.

**Abstraction** T-Abs first checks if the body of the expression has type result $R$ in the typing environment extended with the bound variable $x$ of type $\tau$. $R$ consists of four parts: the return type $\tau$, the true proposition $\psi_+$ which reveals type information about the bound variable $x$ when the return value is non-false, the false proposition $\psi_-$ otherwise, and a symbolic object. Then the lambda is assigned type $\exists X. x{:}\tau \to R$. This rule also shows $X$ might appear in $\tau$ and $R$.

**Application** T-App handles function application. It checks if $e_1$ is a function, and it extends the type environment with $\psi_{1+}$ to ensure the type of $e_2$ is a subtype of the argument type of $e_1$. After doing capture-avoiding substitution of $o_2$ for the occurrences of $x$ in the existential type result $\exists X.R$, it automatically unpacks the type

**Type Checking Extracted Methods**

$$\Delta_s(\mathsf{ctor}(sn, \tau, \overrightarrow{sp\ v_p})) = x{:}\tau \to (sn(\tau, \overrightarrow{sp})\ ;\ \mathbb{TT} \mid \mathbb{FF}\ ;\ \emptyset)$$
$$\Delta_s(\mathsf{acc}(sn(\tau, \overrightarrow{sp}))) = x{:}sn(\tau, \overrightarrow{sp}) \to (\tau\ ;\ \mathbb{TT} \mid \mathbb{TT}\ ;\ \emptyset)$$
$$\Delta_s(\mathsf{pred}(sn(\tau, \overrightarrow{sp}))) = x{:}\top \to (\mathbf{B}\ ;\ x \in sn(\tau, \overrightarrow{sp}) \mid x \notin sn(\tau, \overrightarrow{sp})\ ;\ \emptyset)$$
$$\Delta_s(\mathsf{p\text{-}pred}(sp_i)) = x{:}\top \to (\mathbf{B}\ ;\ x \in \mathbf{Has\text{-}Prop}(sp_i) \mid x \notin \mathbf{Has\text{-}Prop}(sp_i)\ ;\ \emptyset)$$
$$\Delta_s(\mathsf{p\text{-}acc}(sp_i, \tau)) = x{:}\mathbf{Has\text{-}Prop}(sp_i) \to \exists X.(\tau\ ;\ x \in X \mid \mathbb{TT}\ ;\ \emptyset)$$

**Figure 3** Extensions of Meta-function for Typing Rules

result if $X$ appears in $R$. Otherwise, the quantifier is simply ignored. The automatic unpacking is crucial to type checking method extraction. Consider function application (`(custom-write-accessor ins) ins`). The type of `custom-write-accessor` is $x{:}\mathbf{Has\text{-}Prop}(sp_{cw}) \to \exists X_s.((x{:}X_s \to \mathbf{Void}\ ;\ \psi_+ \mid \psi_-\ ;\ o)\ ;\ x \in X_s \mid \mathbb{TT}\ ;\ o')$ and `ins` has type $\mathbf{Has\text{-}Prop}(sp_{cw})$. (`custom-write-accessor ins`) extracts a method from `ins`. The method's type describes that the argument of the previous method extraction is the method receiver, i.e. it has a unique type so that we cannot later apply any value of type $\mathbf{Has\text{-}Prop}(sp_{cw})$ other than `ins`.

**Local bindings**  Our calculus has two new typing rules to check the corresponding new binding forms.

- With the expected type $\tau$, T-Let-Struct-Property creates a named structure property $sp$ along with its property predicate and access procedure. The property descriptor has type $\mathbf{Prop}(\tau)$, where $\tau$ is the expected type for property values. The predicate procedure's type is similar to other type predicates': if the argument passes the predicate, it is of type $\mathbf{Has\text{-}Prop}(sp)$. The accessor's type is built on an existential function type, whose body is type $\tau$ where the receiver type **Self** is replaced with the existential quantifier $X$. Lastly, the rule assigns these three types to three variables and extends the type environment to ensure $e$ is well typed. The final type result has all the bindings erased.
- T-Let-Struct is similar to T-Let-Struct-Property. It creates a structure type with the name $sn$, field type $\tau$, and a collection of property names $\overrightarrow{sp}$ and value expressions $\overrightarrow{e_p}$. Then it checks if the type of each property value match the expected type from the property name. If the latter contains the receiver type **Self**, the type checker will substitute it with the current structure type. The types of the resulting constructor and field accessor procedure are straightforward: the former takes an argument of the field type and returns an instance of the structure, whereas the latter does the opposite. The predicate procedure's type is no different from other type predicates' except for the specific type in the latent propositions. Lastly, the rule assigns these three types to three variables and extends the typing environment to ensure $e$ is well typed. The final type result has all the bindings erased.

**Subsumption and subtyping**  T-Subsume lifts the type result of an expression to a larger one through subtyping rules, which are defined in the usual manner. Our extension adds two new rules: 1. S-Fun arranges the argument types and type results in existential functions the same way as those in normal functions as long as the type

T-Subsume
$$\frac{\Gamma \vdash e : R' \quad \Gamma \vdash R' <: R}{\Gamma \vdash e : R}$$

S-Fun
$$\frac{\Gamma \vdash \tau_2 <: \tau_1 \quad X \mathbin{\#} \Gamma \quad \Gamma, x \in \tau_2 \vdash R_1 <: R_2}{\Gamma \vdash \exists X. x{:}\tau_1 \to R_1 <: \exists X. x{:}\tau_2 \to R_2}$$

S-Struct
$$\frac{\Gamma \vdash sp : (\mathbf{Prop}(\tau)\,;\,\mathbb{TT}\,|\,\mathbb{FF}\,;\,o)}{\Gamma \vdash sn(\tau, \vec{sp}) <: \mathbf{Has\text{-}Prop}(sp)}$$

**Figure 4** Extension of Subtyping Rules

variable appears in the same place 2. S-Struct describes a structure type is a subtype of each well-typed property attached to the structure.

#### 4.2.1 Generativity of let-struct and let-struct-property

In our system, **let-struct** is generative, while **let-struct-property** is not. Consider the following code:

```
(let-struct ((mkfoo foo? foo-a) (foo N []))
 (let ([y (let-struct ([mkfoo^ foo?^ foo-b] (foo B []))
           (mkfoo^ true))])
   (foo-a y)) ;; type error
```

The type checker will report that the domain of `foo-a` does not match the type of `y`, even though the name of the structure type of `y` is `foo`. However, for **let-struct-property**, the name is a fundamental part of the type. If the system allowed re-use of names in structure type properties, then the system would be unsound. Consider the following code:

```
(let-struct-property ((p p? p-acc) (prop N))
  (let-struct ([mkfoo foo? foo-b] (foo N [p 42]))
    (let ([v (mkfoo 10)])
       (let-struct-property ((p1 p?1 p-acc1) (prop x:Self → N))
         ((p-acc1 v) v))))) ;; runtime error
```

We first create a structure type property named `prop`, and attach it to the structure type `foo`. On line 4, we create another property also named `prop` with a different expected property value type, $x{:}\mathbf{Self} \to \mathrm{N}$. When the type checker checks `(p-acc1 v)` on line 5, it only ensures `v` is of a structure type with a property named `prop`. Since the structure type of `v` happens to meet the condition, `(p-acc1 v)` is well typed and so is `((p-acc1 v) v)`. However, at run time the extracted value from `v` will be 42, which is not applicable and will cause a runtime type error. This error, and re-use of structure type property names altogether, is ruled out by the freshness condition in T-Let-Struct-Property.

#### 4.2.2 A Worked Example

To illustrate how all the typing rule extensions help check method extraction, let us work an example:

**Type Checking Extracted Methods**

```
1  (let-struct-property ((pnorm norm? norm-accessor)
2                        (norm x:Self → N))
3    (let-struct ((mkpoint point? point-x)
4                 (point N [(pnorm (λ(this : point(N, pnorm))
5                                    (point-x this)))]))
6      ((λ(v : Has-Prop(norm))
7         ((norm-accessor v) v))
8       (mkpoint 3))))
```

**let-struct-property** creates a structure type property called norm with the expected type $x$:**Self** $\to$ N for property values later supplied in structure definitions. pnorm, norm?, and norm-accessor are bound to the property descriptor, predicate procedure and accessor procedure respectively. pnorm has type **Prop**($x$:**Self** $\to$ N) and norm-accessor has type $x$:**Has-Prop**($norm$) $\to \exists X. ((x{:}X \to N\,;\,\psi_+ \,|\, \psi_- \,;\, o)\,;\, x \in X \,|\, \mathbb{TT}\,;\, o')$. Then we attach the property pnorm to the structure point in its definition. T-Let-struct allows us to get the expected type $x$:**Self** $\to$ N from pnorm's type, substitute the current structure type point for **Self**, and use the result $x$:point(N, pnorm) $\to$ N to successfully check the property value. In the subsequent function application, we first ensure the lambda is well typed. By using T-App, the extracted method from (norm-accessor v) has type $x{:}X \to N$, and the true proposition $x \in X$ from the type result of applying norm-accessor gives $v$ the unique type $X$, therefore the immediate invocation of the extracted method with $v$ is also well typed. Lastly, let us turn to the argument to the lambda on line 8. By S-Struct and T-Subsume, since (mkpoint 3) is a point, a subtype of **Has-Prop**($norm$), it is a valid argument to the lambda.

### 4.3 Semantics, Models and Soundness

**Semantics** Our calculus uses an environment-based big-step reduction semantics described. The core judgment $\rho \vdash e \Downarrow v$ states that expression $e$ evaluates to value $v$ in environment $\rho$, where variables are mapped to closed values. The definition of values are shown in figure 1. See figure 13 in appendix A for a full definition of the evaluation rules.

**Soundness**

**Theorem 1.** *(Type Soundness for $\lambda_{ETR}$). If $\vdash e : \tau$ and $\vdash e \Downarrow v$ then $\vdash v : \tau$*

The theorem states the type safety of a closed-term program in our system with respect to big-step reduction semantics in the usual manner. In particular, the propositions and objects in type results are irrelevant. But in order to help prove the soundness of our calculus, we adopt the full form of the typing judgment in the following lemmas in addition to the same model-theoretic approach from the previous work [13, 17]

**Models** In $\lambda_{ETR}$, a model is any value environment $\rho$. The relation "$\rho$ satisfies $\psi$" is written $\rho \models \psi$, and it states that the proposition $\psi$ holds given the assignment to its free variables in the environment $\rho$. The relation extends to a proposition environment in a point-wise manner. See figure 15 in Appendix A for a full definition of the model relation.

Our first lemma states that our proof theory respects our model.

**Lemma 1.** *If $\rho \models \Gamma$ and $\Gamma \vdash \psi$, then $\rho \models \psi$*

*Proof.* Do structural induction on derivations of $\Gamma \vdash \psi$ □

With Lemma 1 and our operational semantics, we can prove the next lemma crucial to the soundness of our calculus.

**Lemma 2.** *If $\Gamma \vdash e : (\tau \,;\, \psi_+ \,|\, \psi_- \,;\, o)$, $\rho \models \Gamma$ and $\rho \vdash e \Downarrow v$ then all of the following hold:*

1. *$o = \emptyset$ or $\rho(o) = v$*
2. *either $v \neq$ false and $\rho \models \psi_+$, or $v =$ false and $\rho \models \psi_-$*
3. *and $\vdash v : (\tau \,;\, \psi'_+ \,|\, \psi'_- \,;\, o')$ for some $\psi'_+$, $\psi'_-$ and $o'$*

*Proof.* To make the Lemma easier to prove, we slightly modify our typing judgment so that it includes a store to keep track of free type variables. Since the modified system has stronger constraints, our original system is also sound. See appendix B for the complete proof.

Do induction on the derivation of $\rho \vdash e \Downarrow v$. □

We can now easily prove the type soundness of $\lambda_{ETR}$:

*Proof of Theorem 1.* Corollary of Lemma 2. □

Note that our approach does not address diverging or stuck terms, which is the standard drawback of big-step soundness proof. To solve this issue, we could do the following: 1. add a value, *error*, of type $\bot$ 2. add evaluation rules to generate *error* for every stuck terms and add rules to propagate *error* upward 3. prove that the reduction of a well typed term is impossible to be *error*.

## 5 Implementation

In our model, T-Let-Struct is generative but T-Let-Struct-Property is not. However, the corresponding Racket procedures, `make-struct-type-property` and `make-struct-type`, are both generative. We therefore must restrict Typed Racket programs to avoid violating our assumptions. To accomplish this, Typed Racket requires that all `struct` forms and definitions of structure type properties, i.e. (`define-value (pname pred acc) (make-struct-type-property 'name)`) must appear at the top-level of a module, and indexes them with binding information from the definition. This ensures that each such definition is executed only once, and that similarly-named definitions in different modules are kept distinct.

**Type Checking Extracted Methods**

**Binary Methods**   The built-in structure type property `prop:equal+hash` requires its values to contain an equality-checking predicate, which tests if the receiver and second parameter are equal. Inspired by **MyType** in the programming language TOOPL [3], we created **Imp** to denote the implementing structure type so we could annotate the property with (`StructProperty (-> Self Imp (-> Any Any Boolean) Booelean)`). However, how to translate **Imp** in a `StructProperty` type to the type of the corresponding property accessor is still a work in progress, therefore our solution to the binary method problem [5] is incomplete, and **Imp** is not exposed to developers.

**Proposition Propagation**   Since Typed Racket's support for normal function application predates the existential one, our changes that are compatible with the existing code is to handle it parallel to normal function application. The type checker simply needs to use the body of the existential type result to do the rest of type checking in the usual fashion. A difference in our implementation from our model is we do not only use the true latent proposition of the type result of the function, but also propagate it to the lexical proposition environment in order to check subsequent expressions. Consider the following example of using the structure `point2d` defined in listing 3:

```
1 (define p (point2d 10 20))
2 ((custom-write-accessor p) p)
3 (define q (point2d 42 24))
4 (define cw (custom-write-accessor q))
5 (printf "x of q is ~a " (point2d-x q))
6 (cw q)
```

Line 2 shows an example of applying the extracted method to the instance immediately. In this case, extending the typing environment only to check the method application would suffice. However, after extracting a method from an structure instance, developers can manipulate it with the normal structure operations besides applying it to the method, as shown on line 4 – 7. Thus in order to check the following expressions, we need to add the proposition to the typing environment.

On line 6, after extracting `custom-write` from *q*, it is also applied to *point2d-x*. To type check such a program, Typed Racket uses intersection types[1]. Initially, *q* is of type `point2d`. After type checking on line 5, *q* is assigned a unique receiver type $X$, and *q* also keeps its original type, i.e. *q* is of type *point2d*$\wedge X$.

**Contracts**   Interaction between typed code in Typed Racket and untyped code in Racket are protected [10, 11] by contracts [7]. When a typed module exports identifiers to an untyped module, their types are converted to corresponding contracts that ensure the safety of the program at run-time. When a typed module imports identifiers from an untyped module, developers need to annotate them with types that are assumed to be always correct for typed code.

Consider the code defined in listing 4. The enclosing module above is untyped. In its typed submodule, we create the structure type property `foo`, its predicate and accessor procedures. We also define a function that takes a value of any structure type associated with `foo` and returns true. All those identifiers are exported to the enclosing module along with the contracts converted from their types. The contract of type (`Has-Struct-Property prop:foo`) monitors whether a contracted value is an instance of a structure associated with property `prop:foo`. For `foo-ref`, Typed Racket generates a dependent contract for the function. The contract on the return function checks if the invoking argument is an identical value to the receiver. When the check fails, an error is raised.

■ **Listing 4** Type and untyped code interaction

```racket
#lang racket
(module typed typed/racket
  (provide prop:foo foo? dummy)
  (: prop:foo (Struct-Property (-> Self Number)))
  (: foo? (-> Any Boolean : (Has-Struct-Property prop:foo)))
  (: foo-ref (Some (X)
                (-> (Has-Struct-Property prop:foo) (-> X Number) : X)))
  (define-values (prop:foo foo? foo-ref) (make-struct-type-property 'foo))
  (define (dummy [x : (Has-Struct-Property prop:foo)])  : Boolean
     true)
(require 'typed)
(struct world [] #:property prop:foo (lambda (self) 10))
(define x (world))
((foo-ref x) x)
;; raise an exception that the invoking argument is not identical to x
((foo-ref x) (world))
```

Generating contracts from type **Self** is more complex. When bindings are defined in typed modules, those modules are in positive position. For the contract on `prop:foo`, it is provided by the module `typed` to the enclosing module. The contract also specifies that a property value provided by the untyped side should be a function, which makes **Self** in positive position. In this case, typed parts of a program are type checked, therefore the contract for **Self** can be as permissive as possible. On the other hand, when **Self** appears in negative position, i.e. a property name is provided by an untyped module, as shown in the following code, we would fail to gather enough information from the untyped side to create a contract for **Self**:

```racket
#lang racket
(module untyped racket
  (provide prop:foo)
  (define-values (prop:foo foo? foo-ref)
     (make-struct-type-property 'foo)))
(module typed typed/racket
  (require/typed (submod ".." untyped)
    [prop:foo (Struct-Property (-> Self Number))]))
```

## 6 Evaluation

As of Racket 7.9, there were officially 2649 packages on Racket's package catalog, 164 of which were written in or depended on Typed Racket. We divided the evaluation of the impact of our changes into two categories. First, our investigation showed that, 40 of those 164 packages used structure type properties, relying on the previous unsound support. It is also worth pointing out that one typed package, typed-struct-props, provided partial support for structure type properties before our implementation. In

**Type Checking Extracted Methods**

order to ensure that our changes to Typed Racket would not break existing packages, we tested them using our modified version of Typed Racket, including type specifications for all struct type properties provided by the standard library.

When enabling sound checking of struct type properties, only **two** of 40 failed to type check. One failure was simply that an additional type annotation was needed in the property value expression. The other relied on an unsound use of occurrence typing, as shown below:

```
(struct bitmap<%>
  ([convert : (Option (-> Bitmap<%> Symbol Any Any))]
   [shadow : Phantom-Bytes]
   [surface : Bitmap-Surface])
  #:type-name Bitmap<%>
  #:property prop:convertible
  (λ [self mime fallback]
    (with-handlers ([exn? (λ ([e : exn]) (invalid-convert self mime
    ↪  fallback))])
      (cond
        [(bitmap<%>-convert self) => (λ (c) (c self mime fallback))]
        [(bitmap? self) (graphics-convert self mime fallback)]))))
```

The definition of the structure `bitmap<%>` specifies `Bitmap<%>` as the type name for instances of the declared structure. The package developer had assumed the initial type of parameter `self` to be `Any`, therefore a type predicate `bitmap?` was used to refine `self`'s type in the second `cond` clause. Once we enabled type checking on structure property values, `self` would be of type `Bitmap<%>` as declared in this snippet, which has nothing in common with the built-in type `bitmap`.

Our fix was to avoid using the `bitmap?` predicate. Since `bitmap<%>-convert` was not a type predicate, we directly relied on the result of (`bitmap<%>-convert self`):

```
(with-handlers ([exn? (λ [e : exn] (invalid-convert self mime fallback))])
  (define convert (or (bitmap<%>-convert self) graphics-convert))
  (convert self mime fallback))
```

We submitted a pull request to the author of this package, and it was accepted; the revised program works with or without our changes.

Second, we investigated the usage of structures and structure type properties in 2485 untyped packages. We found out that 878 packages defined structures, 243 of which specified a variety of structure type properties via `#:property` in their structure definitions. 45 packages that used structure type properties also created structure type properties through `make-struct-type-property`. In addition, four other packages did not use any structure type properties, but defined and declared properties as exports. Some racket libraries provided functions that require arguments to be instances of structures associated with certain structure type properties. For example, `convert` from the library `file/convert` requires the first argument to be an instance of a `prop:convertible`-attached structure. For `serialize` from the library `racket/serialize`, one type of serializable values are instances of structures with `prop:serializable`. Our investigation showed there were 14 and 19 packages that used `convert` and `serialize` respectively without defining structures with corresponding structure properties. Our high level investigation did not focus on whether structure type properties related code in those packages would be well-typed or even typeable if they were to switch the implementation language from Racket to Typed Racket, but our sound support for structure type properties in Typed Racket will ease the potential transition.

## 7 Related Work

While languages with method extraction have existed for decades, the problem of type checking for these idioms has received little study; we survey the existing literature here.

**Methods as Functions**   Records are used to describe objects and existential types to ensure encapsulation [4]. Here, we keep most of the notation from the work of Pierce and Turner [4]: {f = v ...} is a literal record, {| f : τ ...|} is a record type, and standard pack and unpack operations for existential types. We write r[key] for field access. For example, if variable p is a one-dimensional coordinate value {x=42} of type {| x : Int |}, p[x] access the x field of p and produces 42. With this in hand, consider the following code:

```
Point = Some(X) {|state: X, meth : {|get: X -> Int, set: X -> Int -> X|}|}
p1 = < {|x: Int|}, {state = {x = 5},
                    meth = {set = fun(s:{|x:int|}, i:Int) {x=i},
                            get = fun(s:{|x:int|}) s.x}}> : Point
PointGet(p) = fun(p : Point)
   open p as [X, r] in <r, r[meth][get](r[state])> : Point
 end;
```

Type Point is an alias to an existential type whose body is a record type. p1 is a value of type Point. For p1, the witness type is also a record type. To encode message sending for such an object, explicitly unpacking an existential package is required as shown in the definition of the method PointGet of the object Point.

For method extraction for get, we would want to implement naive encoding in a similar fashion, as shown by NaiveExtractGet defined below.

```
NaiveExtractGet(p) = fun(p : Point)
  (open p as [X, r] in r[meth][get]) : X -> Int
  (* error: X is out of scope*)
end;

ExtractGet(p) = fun(p : Point)
  (open p as [X, r] in fun() r[meth][get](r[state])) : -> Int
end;
```

However, this definition would not work, because the later supplied receiver is of an existential type, while the internal get expects a value of the hidden record type that cannot be used outside the scope. Therefore a general solution is to avoid passing the receiver by closing it over a function, as shown by ExtractGet defined above. Unfortunately, this encoding would not be backward-compatible with how structure type properties are used in Racket.

**Type Checking Extracted Methods**

**Refinement Types**   Refinement types offers another solution to encode method extraction. Consider the following example in Liquid Haskell [16]:

```haskell
data Foo = Foo {i :: Int, get_i :: Foo -> Int}
{-@ extract_get_i :: n: Foo -> {m:Foo | m == n} -> Int @-}
extract_get_i ::  Foo -> Foo -> Int
extract_get_i ins@Foo {i = i, get_i = get_i} = get_i
```

We use a recursive record type to represent a class. The record type `Foo` contains a data field, `i`, and the other field `get_id` as a field accessor function. The function `extract_get_i` takes in a `Foo` instance, and returns its method `get_i`. By specifying the refinement type `m:Foo | m == n`, we enforce the invariance on the self parameter for method extraction, i.e. this extracted method only accepts the same `Foo` instance where it is extracted. Note that refinement typing in Liquid Haskell is supported by an external solver, whereas our approach is a combination of occurrence typing and existential types.

Kent, Kempe, and Tobin-Hochstadt [17] extend Typed Racket with refinement types. We adopt several of their innovations, but they do not include a solver sufficient to handle our use case.

■ **Listing 5**   Node of Singly Linked List
```java
public class Node {
  private int data = 0;
  Node next = null;
  public Node(int d) {
    this.data = d;
  }
  public void setNext(Node next) {
    this.next = next;
  }
}
```

■ **Listing 6**   Node of Doubly Linked List
```java
public class DNode extends Node{
  DNode prev = null;
  public DNode(int d) {
    super(d);
  }
  @Override
  public void setNext(Node next) {
    super.setNext(next);
    ((DNode)next).setPrev(this);
  }
  public void setPrev(DNode prev) {
    this.prev = prev;
  }
}
```

**Self Type**   Bruce [3] proposes a new type called `MyType` in the programming language TOOPL to represent the type of the implementing class so as to avoid dynamic downcasting inside a method. As mentioned in our discussion about binary methods in paragraph 5, this concept serves a similar purpose of $\lambda_{ETR}$'s the implementing structure type **Imp**, and it is different from the receiver type **Self**.

Consider the Java code defined in listings 5 and 6. We define the class `Node` for singly linked list and a subclass, `DNode`, that supports doubly-linked lists. When `DNode`'s `setNext` is invoked, we have to downcast `next` to be a `DNode` before we invoke `setPrev` on `next`. This shows a potential run-time type error when `next` is not an instance of `DNode`. Using **Self** to annotate `next` would eliminate the downcasting, but then the method `setNext` would only be allowed to take the receiver as the argument, making the method useless. If Java adopted `MyType` from TOOPL, the Java code above would become:

```java
1  public class Node {                      public class DNode extends Node{
2    MyType next = null;                      MyType prev = null;
3    /*                                       /* ... */
4       ...                                   @Override
5    */                                       public void setNext(MyType next) {
6    public void setNext(MyType next){          super.setNext(next);
7      this.next = next;                        next.setPrev(this);
8    }                                        }
9  }                                        }
```

In the new implementation, the type of `next` in `setNext` is modified to be `MyType`. Inside the two classes, `MyType` is interpreted as `Node` and `DNode` respectively. This ensures `setPrev` on line 6 of `DNode` can be safely called without dynamic downcasting.

**Existing Languages** Industrial products such as TypeScript, Flow, Hack, Sorbet, as we have shown in the second section, chose to skip sound type checking on method extraction. Java developers can extract methods from a class via reflection API and invoke them with objects and other arguments dynamically. However, type checking in this approach is weak. Developers must rely on exceptions to ensure the run-time safety of programs:

```java
try {
    Method setNext = Node.class.getMethod("setNext", Node.class);
    setNext.invoke(new DNode(42), new Node(10));
} catch (NoSuchMethodException e) {
    err.format("class Node doesn't have a method named setNext");
} catch (IllegalAccessException e) {
    e.printStackTrace();
}
```

In C++ [25], it is possible to create and invoke so-called "pointer-to-member" functions, by using the `std::invoke` operation. However, while this allows supplying a receiver argument, these functions are statically dispatched and do not participate in inheritance-based subtyping. Thus, the programs considered here are either statically rejected or dispatched to a super class method, ignoring the presence of an overriding declaration.

## 8 Conclusion

In this paper, we have described how the integration of occurrence typing and existential types is used to soundly type check method extraction. The combination allows programmers to continue the scripts-to-programs progress by adding strong static guarantee with little or no modification to original code. We have surveyed how existing gradual type systems are unsound in the presence of method extraction. We have also presented a formal model and soundness proof. Our evaluation on the impact of release of the feature of Typed Racket on existing packages shows our design goals have been met. In the future, we aim to build on this success to give types to Racket's generic methods.

**Type Checking Extracted Methods**

**Acknowledgements**   This work was supported by the National Science Foundation, awards 1763922 and 1823244.

**Type Checking Extracted Methods**

## A  Full Formal Model

$$
\begin{array}{rl}
op ::= & \text{not} \mid \text{add1} \mid \text{nat?} \mid ... \\
e ::= & \\
& \mid x \mid sn \mid sp \\
& \mid n \mid \text{true} \mid \text{false} \mid op \\
& \mid \lambda x{:}\tau.e \mid (e\ e) \\
& \mid (\textbf{if}\ e\ e\ e) \\
& \mid (\textbf{let}\ (x\ e)\ e) \\
& \mid (\textbf{let-struct}\ ((x\ x\ x)\ (sn\ \tau\ \overrightarrow{(sp\ e)}))\ e) \\
& \mid (\textbf{let-struct-property}\ ((sp\ x\ x)\ (x\ \tau)))\ e) \\
& \mid (\textbf{cons}\ e\ e) \\
& \mid (\textbf{fst}\ e) \mid (\textbf{snd}\ e) \\
v ::= & \\
& \mid n \mid \text{true} \mid \text{false} \mid op \\
& \mid \langle v, v \rangle \mid [\rho, \lambda x{:}\tau.e] \\
& \mid sn(v : \tau, \overrightarrow{sp\ v}) \\
& \mid \text{pd}(sp) \\
& \mid so \\
so := & \\
& \mid \text{ctor}(x, \tau, \overrightarrow{sp\ v}) \mid \text{pred}(sn(\tau, \overrightarrow{sp})) \mid \text{acc}(sn(\tau, \overrightarrow{sp})) \\
& \mid \text{p-pred}(sp) \mid \text{p-acc}(sp, \tau) \\
\tau, \sigma ::= & \\
& \mid \top \\
& \mid \textbf{N} \mid \textbf{T} \mid \textbf{F} \mid \tau \times \tau \\
& \mid (\bigcup \vec{\tau}) \\
& \mid sn(\tau, \overrightarrow{sp}) \\
& \mid \textbf{Prop}(\tau) \\
& \mid \textbf{Has-Prop}(sp) \\
& \mid \textbf{Self} \\
& \mid X \\
& \mid \exists X.\ x{:}\tau \to R \\
\psi ::= & \\
& \mid \mathbb{TT} \mid \mathbb{FF} \\
& \mid o \in \tau \mid o \notin \tau \\
& \mid \psi \wedge \psi \mid \psi \vee \psi \\
\varphi ::= & \text{fst} \mid \text{snd} \\
o ::= & \\
& \mid \emptyset \\
& \mid x \\
& \mid (\vec{\varphi}\ o) \\
\xi ::= & \psi \mid sp \\
R ::= & (\tau\ ;\ \psi \mid \psi\ ;\ o) \\
\Gamma ::= & \overrightarrow{\xi} \\
\rho ::= & \overrightarrow{x \mapsto \vec{v}}
\end{array}
$$

| | |
|---|---|
| | **Primitive Ops** |
| | **Expressions** |
| | variable |
| | base values |
| | abstraction, application |
| | conditional |
| | local binding |
| | structure binding |
| | structure property binding |
| | pair construction |
| | field access |
| | **Values** |
| | base values |
| | pair, closure |
| | structure instance |
| | struct property descriptor |
| | struct related operations |
| | **Struct-Related-Operations** |
| | ops for structure instance |
| | ops for structure properties |
| | **Types** |
| | universal type |
| | basic types |
| | untagged union type |
| | struct type |
| | struct property type |
| | has struct property type |
| | the receiver type |
| | type variable |
| | existential function type |
| | **Propositions** |
| | trivial/absurd prop |
| | atomic prop |
| | compound props |
| | **Fields** |
| | **Symbolic Objects** |
| | null object |
| | variable reference |
| | object field reference |
| | **Environment Elements** |
| | **Type Result** |
| | **Environments** |
| | **Runtime Environments** |

**Figure 5** Syntax

## Type Checking Extracted Methods

T-Nat
$\Gamma \vdash n : (\mathbf{N}\,;\,\mathbb{TT}\,|\,\mathbb{FF}\,;\,\emptyset)$

T-True
$\Gamma \vdash \text{true} : (\mathbf{T}\,;\,\mathbb{TT}\,|\,\mathbb{FF}\,;\,\emptyset)$

T-False
$\Gamma \vdash \text{false} : (\mathbf{F}\,;\,\mathbb{FF}\,|\,\mathbb{TT}\,;\,\emptyset)$

T-Property-Descriptor
$\Gamma \vdash \text{pd}(sp) : (\mathbf{Prop}(\tau)\,;\,\mathbb{TT}\,|\,\mathbb{FF}\,;\,\emptyset)$

T-Struct-Instance
$\Gamma \vdash sn(v:\tau,\overrightarrow{sp\ v_p}) : (sn(\tau,\overrightarrow{sp})\,;\,\mathbb{TT}\,|\,\mathbb{FF}\,;\,\emptyset)$

T-Struct-Related-Operations
$\Gamma \vdash so : (\Delta_s(so)\,;\,\mathbb{TT}\,|\,\mathbb{FF}\,;\,\emptyset)$

T-Var
$$\frac{\Gamma \vdash x \in \tau}{\Gamma \vdash x : (\tau\,;\,x \notin \mathsf{F}\,|\,x \in \mathsf{F}\,;\,x)}$$

T-Abs
$$\frac{\Gamma, x \in \tau \vdash e : R}{\Gamma \vdash \lambda x{:}\tau.e : (\exists X.\,x{:}\tau \to R\,;\,\mathbb{TT}\,|\,\mathbb{FF}\,;\,\emptyset)}$$

T-Subsume
$$\frac{\Gamma \vdash e : R' \quad \Gamma \vdash R' <: R}{\Gamma \vdash e : R}$$

T-Prim
$\Gamma \vdash op : (\Delta(op)\,;\,\mathbb{TT}\,|\,\mathbb{FF}\,;\,\emptyset)$

T-If
$$\frac{\Gamma \vdash e_1 : (\top\,;\,\psi_{1+}\,|\,\psi_{1-}\,;\,\emptyset) \quad \Gamma, \psi_{1+} \vdash e_2 : R \quad \Gamma, \psi_{1-} \vdash e_3 : R}{\Gamma \vdash (\mathbf{if}\ e_1\ e_2\ e_3) : R}$$

T-Let
$$\frac{\Gamma \vdash e_1 : (\tau_1\,;\,\psi_{1+}\,|\,\psi_{1-}\,;\,o_1) \quad \psi_x = (x \notin \mathsf{F} \wedge \psi_{1+}) \vee (x \in \mathsf{F} \wedge \psi_{1-}) \quad \Gamma, x \in \tau, x \equiv o_1, \psi_x \vdash e : R_2}{\Gamma \vdash (\mathbf{let}\ (x\ e_1)\ e_2) : R_2[x \xmapsto{\tau_1} o_1]}$$

T-Let-Struct-Property
$$\frac{\begin{array}{c}\tau_p = \mathbf{Prop}(\tau) \\ \tau_{pred} = x{:}\top \to (\mathbf{B}\,;\,x \in \mathbf{Has\text{-}Prop}(sp)\,|\,x \notin \mathbf{Has\text{-}Prop}(sp)\,;\,\emptyset) \\ \tau_a = x{:}\mathbf{Has\text{-}Prop}(sp) \to \exists X.\,(\tau[\mathbf{Self} \Longmapsto X]\,;\,x \in X\,|\,\mathbb{TT}\,;\,o_3) \\ \Gamma, sp, x_p \in \tau_p, x_{pred} \in \tau_{pred}, x_{acc} \in \tau_a \vdash e : R \\ X\ \#\ \Gamma \quad X\ \#\ sp\ \#\ \Gamma \end{array}}{\Gamma \vdash (\mathbf{let\text{-}struct\text{-}property}\ ((x_p\ x_{pred}\ x_{acc})\ (sp\ \tau)))\ e) : R[sp \Longmapsto \emptyset][x_{pred} \Longmapsto \emptyset][x_{acc} \Longmapsto \emptyset]}$$

T-Let-Struct
$$\frac{\begin{array}{c}\overrightarrow{\Gamma \vdash sp : (\mathbf{Prop}(\tau_p)\,;\,\mathbb{TT}\,|\,\mathbb{FF}\,;\,\emptyset)} \\ \overrightarrow{\Gamma \vdash e_p : (\tau_p[\mathbf{Self} \Longmapsto sn(\tau,\overrightarrow{sp})]\,;\,\psi_+\,|\,\psi_-\,;\,o_1)} \\ \tau_c n = x{:}\tau \to (sn(\tau,\overrightarrow{sp})\,;\,\mathbb{TT}\,|\,\mathbb{FF}\,;\,\emptyset) \\ \tau_p = x{:}\top \to (\mathbf{B}\,;\,x \in sn(\tau,\overrightarrow{sp})\,|\,x \notin sn(\tau,\overrightarrow{sp})\,;\,\emptyset) \\ \tau_a = x{:}sn(\tau,\overrightarrow{sp}) \to (\tau\,;\,\mathbb{TT}\,|\,\mathbb{FF}\,;\,\emptyset) \\ \Gamma, x_{ctor} \in \tau_c, x_{pred} \in \tau_p, x_{acc} \in \tau_a \vdash e : R \end{array}}{\Gamma \vdash (\mathbf{let\text{-}struct}\ ((x_{ctor}\ x_{pred}\ x_{acc})\ (sn\ \tau\ \overrightarrow{(sp\ e_p)})))\ e) : R[x_{ctor} \Longmapsto \emptyset][x_{pred} \Longmapsto \emptyset][x_{acc} \Longmapsto \emptyset]}$$

**Figure 6** Typing Judgment

T-App
$$\frac{\Gamma \vdash e_1 : (x{:}\tau \to \exists X.R \,;\, \psi_{1+} \mid \psi_{1-} \,;\, \emptyset) \quad \Gamma, \psi_{1+} \vdash e_2 : (\sigma \,;\, \psi_{2+} \mid \psi_{2-} \,;\, o_2) \quad \Gamma \vdash \sigma <: \tau \quad X \,\#\, \sigma \quad X \,\#\, \psi_{1+} \quad X \,\#\, \Gamma}{\Gamma \vdash (e_1\, e_2) : R[x \stackrel{\sigma}{\mapsto} o_2]}$$

T-Cons
$$\frac{\Gamma \vdash e_1 : (\tau_1 \,;\, \mathbb{TT} \mid \mathbb{TT} \,;\, o_1) \quad \Gamma, \vdash e_2 : (\tau_2 \,;\, \mathbb{TT} \mid \mathbb{TT} \,;\, o_2)}{\Gamma \vdash (\textbf{cons}\, e_1\, e_2) : (\tau_1 \times \tau_2 \,;\, \mathbb{TT} \mid \mathbb{TT} \,;\, \emptyset)}$$

T-Fst
$$\frac{\Gamma \vdash e : (\tau_1 \times \tau_2 \,;\, \mathbb{TT} \mid \mathbb{TT} \,;\, o) \quad R = (\tau_1 \,;\, \mathbb{TT} \mid \mathbb{TT} \,;\, (\textsf{fst}\, x))}{\Gamma \vdash (\textbf{fst}\, e) : R[x \stackrel{\tau_1}{\mapsto} o]}$$

T-Snd
$$\frac{\Gamma \vdash e : (\tau_1 \times \tau_2 \,;\, \mathbb{TT} \mid \mathbb{TT} \,;\, o) \quad R = (\tau_2 \,;\, \mathbb{TT} \mid \mathbb{TT} \,;\, (\textsf{snd}\, x))}{\Gamma \vdash (\textbf{snd}\, e) : R[x \stackrel{\tau_2}{\mapsto} o]}$$

■ **Figure 7** Typing Judgment Continued

$$
\begin{aligned}
\Delta_s(\textsf{ctor}(sn, \tau, \overrightarrow{sp\, v_p})) &= x{:}\tau \to (sn(\tau, \overrightarrow{sp}) \,;\, \mathbb{TT} \mid \mathbb{FF} \,;\, \emptyset) \\
\Delta_s(\textsf{acc}(sn(\tau, \overrightarrow{sp}))) &= x{:}sn(\tau, \overrightarrow{sp}) \to (\tau \,;\, \mathbb{TT} \mid \mathbb{TT} \,;\, \emptyset) \\
\Delta_s(\textsf{pred}(sn(\tau, \overrightarrow{sp}))) &= x{:}\top \to (\textbf{B} \,;\, x \in sn(\tau, \overrightarrow{sp}) \mid x \notin sn(\tau, \overrightarrow{sp}) \,;\, \emptyset) \\
\Delta_s(\textsf{p-pred}(sp_i)) &= x{:}\top \to (\textbf{B} \,;\, x \in \textbf{Has-Prop}(sp_i) \mid x \notin \textbf{Has-Prop}(sp_i) \,;\, \emptyset) \\
\Delta_s(\textsf{p-acc}(sp_i, \tau)) &= x{:}\textbf{Has-Prop}(sp_i) \to \exists X. (\tau \,;\, x \in X \mid \mathbb{TT} \,;\, \emptyset)
\end{aligned}
$$

■ **Figure 8** Types of Operations on Struct-related Values

$$
\begin{aligned}
\Delta(\textsf{not}) &= x{:}\top \to (\textbf{B} \,;\, x \in \textsf{F} \mid x \notin \textsf{F} \,;\, \emptyset) \\
\Delta(\textsf{add1}) &= x{:}\textbf{N} \to (\textbf{N} \,;\, \mathbb{TT} \mid \mathbb{FF} \,;\, \emptyset) \\
\Delta(\textsf{nat?}) &= x{:}\top \to (\textbf{B} \,;\, x \in \textbf{N} \mid x \notin \textbf{N} \,;\, \emptyset) \\
\Delta(\textsf{bool?}) &= x{:}\top \to (\textbf{B} \,;\, x \in \textbf{B} \mid x \notin \textbf{B} \,;\, \emptyset) \\
\Delta(\textsf{pair?}) &= x{:}\top \to (\textbf{B} \,;\, x \in \top \times \top \mid x \notin \top \times \top \,;\, \emptyset)
\end{aligned}
$$

■ **Figure 9** Types of Primitive Operations

**Base Values** T-Nat shows any natural number has type **N**, and since an **N** is a non-false value, its true proposition is $\mathbb{TT}$ with the false proposition being $\mathbb{FF}$. Rules for other non-false values such as T-True follow a similar specification, while T-False is different. The value is false, therefore its propositions are the opposite of those of non-false values. T-Prim assigns function types to primitives by referring to the metafunction $\Delta$ described in figure 9. Since there is no object referencing the portion of the runtime environment, the object parts of those rules are $\emptyset$.

**Structure Related Values** T-Property-Descriptor shows the named property $\textsf{pd}(sp)$ has type $\textbf{Prop}(\tau)$. T-Struct-Instance shows the structure instance $sn(v : \tau, \overrightarrow{sp\, v_p})$

**Type Checking Extracted Methods**

has type $sn(\tau, \overrightarrow{sp})$. Through the metafunction $\Delta_s$, T-STRUCT-RELATED-OPERATIONS assigns function types to primitive operations for structure instances and structure type properties. Figure 8 describes the definition of $\Delta_s$.

**Variable**  T-VAR assigns type $\tau$ to variable $x$ if the proof system can show that $x$ has type $\tau$. The true and false propositions reflect the two groups of values $x$ referred to in the runtime environment: non-false values and false. The object part follows the definition of the judgment.

**Conditionals**  In T-IF, the condition expression $e_1$ is first checked. If it is well typed, the resulting true proposition and false proposition are used to extend the typing environment to ensure that the two branch expressions $e_2$ and $e_3$ are also well-typed respectively. The true proposition in the final type result is a disjoint union of those from the type results of $e_2$ and $e_3$, and so is the false proposition.

**Local Bindings**

- T-LET first checks if $e_1$ is well-typed, and assign the type of $e_1$ to x. $\psi_x$ accounts for one of the following cases: if $x$ refers to a non-false value, then $\psi_+$ holds; Otherwise, $\psi_-$ holds. Then the rule extends the typing environment with $x$'s type, $\psi_x$, the equivalence relation between $o_1$ and $x$ to ensure $e_2$ is well-typed. The final type result is that of $e_2$ with substitution of $o_1$ for $x$.
- With the expected type $\tau$, T-LET-STRUCT-PROPERTY creates a named structure property $sp$ along with its property predicate and access procedure. The property descriptor has type **Prop**$(\tau)$, where $\tau$ is the expected type for property values. The predicate procedure's type is similar to other type predicates': if the argument passes the predicate, it is of type **Has-Prop**$(sp)$. The accessor's type is built on an existential function type, whose body is type $\tau$ where the receiver type **Self** is replaced with the existential quantifier $X$. Lastly, the rule assigns these three types to three variables and extends the type environment to ensure $e$ is well typed. The final type result has all the bindings erased.
- T-LET-STRUCT is similar to T-LET-STRUCT-PROPERTY. It creates a structure type with the name $sn$, field type $\tau$, and a collection of property names $\overrightarrow{sp}$ and value expressions $\overrightarrow{e_p}$. Then it checks if the type of each property value match the expected type from the property name. If the latter contains the receiver type **Self**, the type checker will substitute it with the current structure type. The types of the resulting constructor and field accessor procedure are straightforward: the former takes an argument of the field type and returns an instance of the structure, whereas the latter does the opposite. The predicate procedure's type is no different from other type predicates' except for the specific type in the latent propositions. Lastly, the rule assigns these three types to three variables and extends the typing environment to ensure $e$ is well typed. The final type result has all the bindings erased.

**Abstraction**  T-ABS first checks if the body of the expression has type result $R$ in the typing environment extended with the bound variable $x$ of type $\tau$. $R$ consists of four parts: the return type $\tau$, the true proposition $\psi_+$ which reveals type information about the bound variable $x$ when the return value is non-false, the false proposition $\psi_-$ otherwise, and a symbolic object. Then the lambda is assigned type $\exists X. x{:}\tau \to R$. This rule also shows $X$ might appear in $\tau$ and $R$.

**Application** T-App handles function application. It checks if $e_1$ is a function, and it extends the type environment with $\psi_{1+}$ to ensure the type of $e_2$ is a subtype of the argument type of $e_1$. After doing capture-avoiding substitution of $o_2$ for the occurrences of $x$ in the existential type result $\exists X. R$, it automatically unpacks the type result if $X$ appears in $R$. Otherwise, the quantifier is simply ignored. The automatic unpacking is crucial to type checking method extraction. Consider function application ((custom-write-accessor ins) ins). The type of custom-write-accessor is $x{:}\textbf{Has-Prop}(sp_{cw}) \to \exists X_s.\,((x{:}X_s \to \textbf{Void}\,;\,\psi_+ \mid \psi_-\,;\,o)\,;\,x \in X_s \mid \mathbb{TT}\,;\,o')$ and ins has type $\textbf{Has-Prop}(sp_{cw})$. (custom-write-accessor ins) extracts a method from ins. The method's type describes that the argument of the previous method extraction is the method receiver, i.e. it has a unique type so that we cannot later apply any value of type $\textbf{Has-Prop}(sp_{cw})$ other than ins.

**Pairs** T-Cons introduces a pair type by ensuring its two components are well typed. T-Fst and T-Snd eliminate a pair type. If an argument is a pair, they include the first and second argument type in their the final type results respectively in addition to prepending an extra path to the symbolic object in the type result of the argument.

**Subsumption and Subtyping** T-Subsume lifts the type result of an expression to a larger one through subtyping rules, which are defined in the usual manner. Our extension adds two new rules: 1. S-Fun arranges the argument types and type results in existential functions the same way as those in normal functions as long as the type variable appears in the same place 2. S-Struct describes a structure type is a subtype of each well-typed property attached to the structure.

**Proof System** Figure 11 describes the logic rules for our calculus. They are directly inherited from $\lambda_{TR}$ with modifications. The first eight rules are introduction and elimination forms that resemble their counterpart in propositional logic. L-Sub says if a typing environment proves an object has a subtype of a larger type, then it also proves the object has the larger type. In L-Not, if a typing environment is incompatible with an object's type, then we can conclude that the object doesn't have the type. L-Bot, serving as "ex falso quodlibet" of sorts in our system, allows us to derive any conclusion if an object has type empty. By L-Update+ and L-Update-, we are able to use multiple positive and negative type statements on an object to refine its type. The refinement is done through the metafunction described in figure 12. Roughly speaking, the metafunction updates the type of some field of an object by doing a conservative intersection of two types when it has the knowledge of the field's type, while computing their difference when it knows the field doesn't have the type.

$\lambda_{ETR}$ also extends the reduction rules of $\lambda_{TR}$ with three rules for structures and structure type properties. B-Let-Struct-Property extends the environment with a

SO-Null
$\Gamma \vdash o <: \emptyset$

S-Refl
$\Gamma \vdash \tau <: \tau$

S-Top
$\Gamma \vdash \tau <: \top$

S-Union-Sub
$\dfrac{\forall \tau \text{ in } \vec{\tau}.\ \Gamma \vdash \tau <: \sigma}{\Gamma \vdash (\bigcup \vec{\tau}) <: \sigma}$

S-Pair
$\dfrac{\Gamma \vdash \tau_1 <: \tau_2 \quad \Gamma \vdash \sigma_1 <: \sigma_2}{\Gamma \vdash \tau_1 \times \sigma_1 <: \tau_2 \times \sigma_2}$

S-Fun
$\dfrac{\Gamma \vdash \tau_2 <: \tau_1 \quad X \# \Gamma \quad \Gamma, x \in \tau_2 \vdash R_1 <: R_2}{\Gamma \vdash \exists X. x{:}\tau_1 \to R_1 <: \exists X. x{:}\tau_2 \to R_2}$

S-Union-Super
$\dfrac{\exists \sigma \text{ in } \vec{\sigma}.\ \Gamma \vdash \tau <: \sigma}{\Gamma \vdash \tau <: (\bigcup \vec{\sigma})}$

S-Result
$\dfrac{\Gamma \vdash \tau_1 <: \tau_2 \quad \Gamma, \psi_{1+} \vdash \psi_{2+} \quad \Gamma \vdash o_1 <: o_2 \quad \Gamma, \psi_{1-} \vdash \psi_{2-}}{\Gamma \vdash (\tau_1\,;\,\psi_{1+}\,|\,\psi_{1-}\,;\,o_1) <: (\tau_2\,;\,\psi_{2+}\,|\,\psi_{2-}\,;\,o_2)}$

S-Struct
$\dfrac{\Gamma \vdash sp\ :\ (\mathbf{Prop}(\tau)\,;\,\mathbb{TT}\,|\,\mathbb{FF}\,;\,o)}{\Gamma \vdash sn(\tau, \vec{sp}) <: \mathbf{Has\text{-}Prop}(sp)}$

**Figure 10** Subtyping

L-Atom
$\dfrac{\psi \in \Gamma}{\Gamma \vdash \psi}$

L-Trivial
$\Gamma \vdash \mathbb{TT}$

L-Absurd
$\dfrac{\Gamma \vdash \mathbb{FF}}{\Gamma \vdash \psi}$

L-AndI
$\dfrac{\Gamma \vdash \psi_1 \quad \Gamma \vdash \psi_2}{\Gamma \vdash \psi_1 \wedge \psi_2}$

L-AndE1
$\dfrac{\Gamma \vdash \psi_1 \wedge \psi_2}{\Gamma \vdash \psi_1}$

L-AndE2
$\dfrac{\Gamma \vdash \psi_1 \wedge \psi_2}{\Gamma \vdash \psi_2}$

L-OrI
$\dfrac{\Gamma \vdash \psi_1 \text{ or } \Gamma \vdash \psi_2}{\Gamma \vdash \psi_1 \vee \psi_2}$

L-OrE
$\dfrac{\Gamma \vdash \psi_1 \vee \psi_2 \quad \Gamma, \psi_1 \vdash \psi \quad \text{or} \quad \Gamma, \psi_2 \vdash \psi}{\Gamma \vdash \psi}$

L-Sub
$\dfrac{\Gamma \vdash o \in \sigma \quad \Gamma \vdash \sigma <: \tau}{\Gamma \vdash o \in \tau}$

L-Not
$\dfrac{\Gamma, o \in \tau \vdash \mathbb{FF}}{\Gamma \vdash o \notin \tau}$

L-Bot
$\dfrac{\Gamma \vdash o \in \bot}{\Gamma \vdash \psi}$

L-Update+
$\dfrac{\Gamma \vdash o \in \tau \quad \Gamma \vdash (\vec{\varphi}\ o) \in \sigma}{\Gamma \vdash o \in \mathsf{update}^+_\Gamma(\tau, \vec{\varphi}, \sigma)}$

L-Update−
$\dfrac{\Gamma \vdash o \in \tau \quad \Gamma \vdash (\vec{\varphi}\ o) \notin \sigma}{\Gamma \vdash o \in \mathsf{update}^-_\Gamma(\tau, \vec{\varphi}, \sigma)}$

**Figure 11** Proof system

property descriptor, its accessor procedure and predicate procedure to evaluate the body. B-Let-Struct evaluates the body in the same way after it gets the values of the property expressions. B-Struct-Related-Operations describes function application of those generated procedures for structure instances. Figure 14 details the metafunction $\delta_s$.

$$\begin{aligned}
\text{update}_\Gamma^\pm(\tau_1 \times \tau_2, \vec{\varphi}::\text{fst}, \sigma) &= \text{update}_\Gamma^\pm(\tau_1, \vec{\varphi}, \sigma) \times \tau_2 \\
\text{update}_\Gamma^\pm(\tau_1 \times \tau_2, \vec{\varphi}::\text{snd}, \sigma) &= \tau_1 \times \text{update}_\Gamma^\pm(\tau_2, \vec{\varphi}, \sigma) \\
\text{update}_\Gamma^+(\tau, \epsilon, \sigma) &= \text{restrict}_\Gamma(\tau, \sigma) \\
\text{update}_\Gamma^-(\tau, \epsilon, \sigma) &= \text{remove}_\Gamma(\tau, \sigma) \\
\text{update}_\Gamma^\pm((\bigcup \vec{\tau}), \vec{\varphi}, \sigma) &= (\bigcup \overrightarrow{\text{update}_\Gamma^\pm(\tau, \vec{\varphi}, \sigma)})
\end{aligned}$$

$$\begin{aligned}
\text{restrict}_\Gamma(\tau, \sigma) &= \bot \quad \text{if } \tau \cap \sigma = \varnothing \\
\text{restrict}_\Gamma((\bigcup \vec{\tau}), \sigma) &= (\bigcup \overrightarrow{\text{restrict}_\Gamma(\tau, \sigma)}) \\
\text{restrict}_\Gamma(\tau, \sigma) &= \tau \quad \text{if } \Gamma \vdash \tau <: \sigma \\
\text{restrict}_\Gamma(\tau, \sigma) &= \sigma \quad \text{otherwise}
\end{aligned}$$

$$\begin{aligned}
\text{remove}_\Gamma(\tau, \sigma) &= \bot \quad \text{if } \Gamma \vdash \tau <: \sigma \\
\text{remove}_\Gamma((\bigcup \vec{\tau}), \sigma) &= (\bigcup \overrightarrow{\text{remove}_\Gamma(\tau, \sigma)}) \\
\text{remove}_\Gamma(\tau, \sigma) &= \tau \quad \text{otherwise}
\end{aligned}$$

**Figure 12** Metafunction Update

## Type Checking Extracted Methods

B-VAL
$$\rho \vdash v \Downarrow v$$

B-VAR
$$\frac{\rho(x) = v}{\rho \vdash x \Downarrow v}$$

B-LET
$$\frac{\rho \vdash e_1 \Downarrow v_1 \quad \rho[x := v_1] \vdash e_2 \Downarrow v}{\rho \vdash (\textbf{let}\,(x\,e_1)\,e_2) \Downarrow v}$$

B-ABS
$$\rho \vdash \lambda x{:}\tau.e \Downarrow [\rho, \lambda x{:}\tau.e]$$

B-FST
$$\frac{\rho \vdash e \Downarrow \langle v_1, v_2 \rangle}{\rho \vdash (\textsf{fst}\,e) \Downarrow v_1}$$

B-SND
$$\frac{\rho \vdash e \Downarrow \langle v_1, v_2 \rangle}{\rho \vdash (\textsf{snd}\,e) \Downarrow v_2}$$

B-BETA
$$\frac{\rho \vdash e_1 \Downarrow [\rho_c, \lambda x{:}\tau.e] \quad \rho \vdash e_2 \Downarrow v_2 \quad \rho_c[x := v_2] \vdash e \Downarrow v}{\rho \vdash (e_1\,e_2) \Downarrow v}$$

B-PRIM
$$\frac{\rho \vdash e_1 \Downarrow op \quad \rho \vdash e_2 \Downarrow v_2 \quad \delta(op, v_2) = v}{\rho \vdash (e_1\,e_2) \Downarrow v}$$

B-IFTRUE
$$\frac{\rho \vdash e_1 \Downarrow v_1 \quad v_1 \neq \textsf{false} \quad \rho \vdash e_2 \Downarrow v}{\rho \vdash (\textbf{if}\,e_1\,e_2\,e_3) \Downarrow v}$$

B-IFFALSE
$$\frac{\rho \vdash e_1 \Downarrow \textsf{false} \quad \rho \vdash e_3 \Downarrow v}{\rho \vdash (\textbf{if}\,e_1\,e_2\,e_3) \Downarrow v}$$

B-PAIR
$$\frac{\rho \vdash e_1 \Downarrow v_1 \quad \rho \vdash e_2 \Downarrow v_2}{\rho \vdash (\textbf{cons}\,e_1\,e_2) \Downarrow \langle v_1, v_2 \rangle}$$

B-LET-STRUCT
$$\frac{\begin{array}{c}\overrightarrow{\rho \vdash e_p \Downarrow v_p} \\ v_{ctor} = \textsf{ctor}(sn, \tau, \overrightarrow{x_p\,v_p}) \\ v_{pred} = \textsf{pred}(sn) \\ v_{acc} = \textsf{acc}(sn) \\ \rho[x_{ctor} := v_{ctor}][x_{pred} := v_{pred}][x_{acc} := v_{acc}] \vdash e \Downarrow v\end{array}}{\rho \vdash (\textbf{let-struct}\,((x_{ctor}\,x_{pred}\,x_{acc})\,(sn\,\tau_f\,\overrightarrow{(x_p\,e_p)}))\,e) \Downarrow v}$$

B-LET-STRUCT-PROPERTY
$$\frac{\begin{array}{c} v_p = \textsf{pd}(x_p) \\ v_{pred} = \textsf{p-pred}(x_p) \\ v_{acc} = \textsf{p-acc}(x_p, \tau) \\ \rho[sp := v_p][x_{pred} := v_{pred}][x_{acc} := v_{acc}] \vdash e \Downarrow v\end{array}}{\rho \vdash (\textbf{let-struct-property}\,((sp\,x_{pred}\,x_{acc})\,(x_p\,\tau)))\,e) \Downarrow v}$$

B-STRUCT-RELATED-OPERATIONS
$$\frac{\rho \vdash e_1 \Downarrow so \quad \rho \vdash e_2 \Downarrow v_1 \quad \delta_s(so, v_1) = v_2}{\rho \vdash (e_1\,e_2) \Downarrow v_2}$$

**Figure 13** Big-step Reduction

$$\begin{aligned}
\delta_s(\mathsf{ctor}(sn, \tau, \overrightarrow{x_p\, v_p}), v) &= sn(v : \tau, \overrightarrow{x_p\, v_p}) \\
\delta_s(\mathsf{acc}(sn(\tau, \overrightarrow{sp})), sn(v : \tau, \overrightarrow{x_p\, v_p})) &= v \\
\delta_s(\mathsf{pred}(sn(\tau, \overrightarrow{sp})), sn(v : \tau, \overrightarrow{x_p\, v_p})) &= true \\
\delta_s(\mathsf{pred}(sn(\tau, \overrightarrow{sp})), v) &= \mathsf{false} \qquad \text{if } v \neq sn(v : \tau, \overrightarrow{x_p\, v_p}) \\
\delta_s(\mathsf{p\text{-}pred}(x_{p_i}), sn(v : \tau, \overrightarrow{x_p\, v_p})) &= \mathsf{true} \\
\delta_s(\mathsf{p\text{-}pred}(x_{p_i}), v) &= \mathsf{false} \qquad \text{if } v \neq sn(v : \tau, \overrightarrow{x_p\, v_p}) \\
\delta_s(\mathsf{p\text{-}acc}(x_{p_i}, \tau), sn(v : \tau, \overrightarrow{x_p\, v_p})) &= v_{p_i}
\end{aligned}$$

**Figure 14** Operations on struct-related values

$$\begin{array}{ccc}
\text{M-Top} & \text{M-Or} & \text{M-And} \\
& & \rho \models \psi_1 \quad \rho \models \psi_2 \\
\rho \models \mathbb{TT} & \dfrac{\rho \models \psi_1 \text{ or } \rho \models \psi_2}{\rho \models \psi_1 \vee \psi_2} & \dfrac{\text{Type variables are distinct in } \psi_1 \text{ and } \psi_2}{\rho \models \psi_1 \wedge \psi_2}
\end{array}$$

$$\begin{array}{ccc}
\text{M-Type} & \text{M-TypeNot} & \text{M-Alias} \\
\dfrac{\vdash \rho(o) : \tau}{\rho \models o \in \tau} & \dfrac{\vdash \rho(o) : \sigma \quad \sigma \cap \tau = \varnothing}{\rho \models o \notin \tau} & \dfrac{\rho(o_1) = \rho(o_2)}{\rho \models o_1 \equiv o_2}
\end{array}$$

**Figure 15** Satisfaction Relation

$$\begin{aligned}
\delta(\mathsf{not}, \mathsf{false}) &= \mathsf{true} \\
\delta(\mathsf{not}, v) &= \mathsf{false} \\
\delta(\mathsf{add1}, n) &= n + 1 \\
\delta(\mathsf{nat?}, n) &= \mathsf{true} \\
\delta(\mathsf{nat?}, v) &= \mathsf{false} \\
\delta(\mathsf{bool?}, \mathsf{true}) &= \mathsf{true} \\
\delta(\mathsf{bool?}, \mathsf{false}) &= \mathsf{true} \\
\delta(\mathsf{bool?}, v) &= \mathsf{false} \\
\delta(\mathsf{pair?}, \langle v, v \rangle) &= \mathsf{true} \\
\delta(\mathsf{pair?}, v) &= \mathsf{false}
\end{aligned}$$

**Figure 16** Primitives

## B  Full Proof for Soundness

To prove the soundness of our calculus, we add to our typing judgement a store to track free type variables: $\Gamma \vdash e : R \mid \overrightarrow{T}$

**Lemma 1.** *If $\rho \models \Gamma$ and $\Gamma \vdash \psi$, then $\rho \models \psi$*

*Proof.* Do structural induction on derivations of $\Gamma \vdash \psi$:

**L-Trivial**  By M-TOP, $\rho \models \mathbb{TT}$.

**L-Atom**  since $\psi \in \Gamma$, $\rho \models \psi$ by assumption.

**L-Absurd**  since $\rho \not\models \mathbb{FF}$, this case is impossible to prove

**L-AndI**  By IH, $\rho \models \psi_1$ and $\rho \models \psi_2$. By M-AND, $\rho \models \psi_1 \wedge \psi_2$

**L-AndE1 and L-AndE2**  By IH, $\rho \models \psi_1 \wedge \psi_2$. By inversion on it, $\rho \models \psi_1$ and $\rho \models \psi_2$

**L-ORI**  By IH, $\rho \models \psi_1$ or $\rho \models \psi_2$. By M-OR, $\rho \models \psi_1 \vee \psi_2$

**L-ORE**  By IH, $\rho \models \psi_1$ or $\rho \models \psi_2$. Since $\rho \models \Gamma$, $\rho \models \Gamma, \psi_1$ or $\rho \models \Gamma, \psi_2$ and so $\rho \models \psi$

**etc...**

□

**Lemma 2.** *If $\Gamma \vdash e : (\tau\,;\,\psi_+ \mid \psi_-\,;\,o) \mid \overrightarrow{T}$, $\rho \models \Gamma$ and $\rho \vdash e \Downarrow v$ then all of the following hold:*

1. *$o = \emptyset$ or $\rho(o) = v$*
2. *either $v \neq$ false and $\rho \models \Gamma, \psi_+$, or $v =$ false and $\rho \models \Gamma, \psi_-$*
3. *and $\vdash v : (\tau\,;\,\psi'_+ \mid \psi'_-\,;\,o') \mid \overrightarrow{T}$ for some $\psi'_+$, $\psi'_-$ and $o'$*

*Proof.* We are applying induction on the derivation of $\rho \vdash e \Downarrow v$. Since the corresponding typing derivation for each evaluation rule can have the non-subsumption rule and T-SUBSUME as its last two rules. To simplify the following proof by cases, we first prove the lemma holds for T-SUBSUME and evaluation derivations if it holds for non-subsumption rules:

By inversion on T-SUBSUME, $\Gamma \vdash e : (\sigma\,;\,\psi'_+ \mid \psi'_-\,;\,o_x) \mid \overrightarrow{T}$, $\Gamma \vdash (\sigma\,;\,\psi'_+ \mid \psi'_-\,;\,x) <: (\tau\,;\,\psi_+ \mid \psi_-\,;\,o)$. Assume our lemma holds for $\Gamma \vdash e : (\sigma\,;\,\psi'_+ \mid \psi'_-\,;\,e_x) \mid \overrightarrow{T}$, Then we are able to prove:

1. $o = \emptyset$. Otherwise, $o = o_x$. Then by IH, $\rho(o_x) = v$ and so $\rho(o) = v$
2. We need to show $\rho \models \Gamma, \psi_+$, if $v \neq$ false: By IH $\rho \models \Gamma, \psi'_+$. By inversion on S-RESULT: $\Gamma, \psi'_+ \vdash \psi_+$. By Lemma 3, $\Gamma, \psi'_+ \vdash \Gamma, \psi_+$. Then by Lemma 1, $\rho \models \Gamma, \psi_+$. For proving $\rho \models \Gamma, \psi_-$ if $v =$ false, the reasoning is similar.
3. By IH, $\Gamma \vdash v : (\sigma\,;\,\psi'_+ \mid \psi'_-\,;\,x) \mid \overrightarrow{T}$. Then $\Gamma \vdash v : (\tau\,;\,\psi_+ \mid \psi_-\,;\,o) \mid \overrightarrow{T}$

Now proceed by cases from induction on the derivation of $\rho \vdash e \Downarrow v$ regardless of T-SUBSUME:

- **B-Val** $\rho \vdash v \Downarrow v$ Do structral induction on v:
  - **Case v = n**: Two rules can be derived as the last one in the typing derivation: T-NAT,T-IEXI. Proceed by cases.

T-Nat $\qquad$ T-True
$\Gamma \vdash n : (\mathbf{N}\,;\,\mathbb{TT}\,|\,\mathbb{FF}\,;\,\emptyset)\,|\,\square \qquad \Gamma \vdash \mathsf{true} : (\mathsf{T}\,;\,\mathbb{TT}\,|\,\mathbb{FF}\,;\,\emptyset)\,|\,\square$

T-False $\qquad\qquad\qquad\qquad$ T-Property-Descriptor
$\Gamma \vdash \mathsf{false} : (\mathsf{F}\,;\,\mathbb{FF}\,|\,\mathbb{TT}\,;\,\emptyset)\,|\,\square \qquad \Gamma \vdash \mathsf{pd}(sp) : (\mathbf{Prop}(\tau)\,;\,\mathbb{TT}\,|\,\mathbb{FF}\,;\,\emptyset)\,|\,\square$

T-Struct-Instance $\qquad\qquad\qquad\qquad$ T-Struct-Related-Operations
$\Gamma \vdash sn(v:\tau, \overrightarrow{sp\ v_p}) : (sn(\tau, \overrightarrow{sp})\,;\,\mathbb{TT}\,|\,\mathbb{FF}\,;\,\emptyset)\,|\,\square \qquad \Gamma \vdash so : (\Delta_s(so)\,;\,\mathbb{TT}\,|\,\mathbb{FF}\,;\,\emptyset)\,|\,\square$

T-Var $\qquad\qquad\qquad\qquad$ T-Abs
$$\dfrac{\Gamma \vdash x \in \tau}{\Gamma \vdash x : (\tau\,;\,x \notin \mathsf{F}\,|\,x \in \mathsf{F}\,;\,x)\,|\,\square} \qquad \dfrac{\Gamma, x \in \tau \vdash e : R\,|\,\overrightarrow{T}}{\Gamma \vdash \lambda x{:}\tau.e : (\exists X.\,x{:}\tau \to R\,;\,\mathbb{TT}\,|\,\mathbb{FF}\,;\,\emptyset)\,|\,\overrightarrow{T}}$$

T-Subsume
$$\dfrac{\Gamma \vdash e : R'\,|\,\overrightarrow{T} \qquad \Gamma \vdash R' <: R}{\Gamma \vdash e : R\,|\,\overrightarrow{T}} \qquad \text{T-Prim}\ \ \Gamma \vdash op : (\Delta(op)\,;\,\mathbb{TT}\,|\,\mathbb{FF}\,;\,\emptyset)\,|\,\square$$

T-If $\qquad\qquad\qquad\qquad$ T-Let
$$\dfrac{\begin{array}{c}\Gamma \vdash e_1 : (\top\,;\,\psi_{1+}\,|\,\psi_{1-}\,;\,\emptyset)\,|\,\overrightarrow{T_1} \\ \Gamma, \overrightarrow{T_1}, \psi_{1+} \vdash e_2 : R\,|\,\overrightarrow{T_2} \\ \Gamma, \overrightarrow{T_1}, \psi_{1-} \vdash e_3 : R\,|\,\overrightarrow{T_3}\end{array}}{\Gamma \vdash (\mathbf{if}\ e_1\ e_2\ e_3) : R\,|\,\overrightarrow{T_1} + \overrightarrow{T_2} + \overrightarrow{T_3}} \qquad \dfrac{\begin{array}{c}\Gamma \vdash e_1 : (\tau_1\,;\,\psi_{1+}\,|\,\psi_{1-}\,;\,o_1)\,|\,\overrightarrow{T_1} \\ \psi_x = (x \notin \mathsf{F} \wedge \psi_{1+}) \vee (x \in \mathsf{F} \wedge \psi_{1-}) \\ \Gamma, \overrightarrow{T_1}, x \in \tau, x \equiv o_1, \psi_x \vdash e : R_2\,|\,\overrightarrow{T_2}\end{array}}{\Gamma \vdash (\mathbf{let}\ (x\ e_1)\ e_2) : R_2[x \xmapsto{\tau_1} o_1]\,|\,\overrightarrow{T_1} + \overrightarrow{T_2}}$$

T-Let-Struct-Property
$$\begin{array}{c}\tau_p = \mathbf{Prop}(\tau) \\ \tau_{pred} = x{:}\top \to (\mathbf{B}\,;\,x \in \mathbf{Has\text{-}Prop}(sp)\,|\,x \notin \mathbf{Has\text{-}Prop}(sp)\,;\,\emptyset) \\ \tau_a = x{:}\mathbf{Has\text{-}Prop}(sp) \to \exists X.\,(\tau[\mathbf{Self} \Mapsto X]\,;\,x \in X\,|\,\mathbb{TT}\,;\,o_3) \\ \Gamma, sp, x_p \in \tau_p, x_{pred} \in \tau_{pred}, x_{acc} \in \tau_a \vdash e : R\,|\,\overrightarrow{T} \\ X\ \#\ \Gamma\ \ X\ \#\ sp\ \#\ \Gamma\end{array}$$
$$\overline{\Gamma \vdash (\mathbf{let\text{-}struct\text{-}property}\ ((x_p\ x_{pred}\ x_{acc})\ (sp\ \tau)))\ e) : R[sp \Mapsto \emptyset][x_{pred} \Mapsto \emptyset][x_{acc} \Mapsto \emptyset]\,|\,\overrightarrow{T}}$$

T-Let-Struct
$$\begin{array}{c}\overrightarrow{\Gamma \vdash sp : (\mathbf{Prop}(\tau_p)\,;\,\mathbb{TT}\,|\,\mathbb{FF}\,;\,\emptyset)\,|\,\square} \\ \overrightarrow{\Gamma \vdash e_p : (\tau_p[\mathbf{Self} \Mapsto sn(\tau, \overrightarrow{sp})]\,;\,\psi_+\,|\,\psi_-\,;\,o_1)\,|\,\square} \\ \tau_c n = x{:}\tau \to (sn(\tau, \overrightarrow{sp})\,;\,\mathbb{TT}\,|\,\mathbb{FF}\,;\,\emptyset) \\ \tau_p = x{:}\top \to (\mathbf{B}\,;\,x \in sn(\tau, \overrightarrow{sp})\,|\,x \notin sn(\tau, \overrightarrow{sp})\,;\,\emptyset) \\ \tau_a = x{:}sn(\tau, \overrightarrow{sp}) \to (\tau\,;\,\mathbb{TT}\,|\,\mathbb{FF}\,;\,\emptyset) \\ \Gamma, x_{ctor} \in \tau_c, x_{pred} \in \tau_p, x_{acc} \in \tau_a \vdash e : R\,|\,\square\end{array}$$
$$\overline{\Gamma \vdash (\mathbf{let\text{-}struct}\ ((x_{ctor}\ x_{pred}\ x_{acc})\ (sn\ \tau\ \overrightarrow{(sp\ e_p)}))\ e) : R[x_{ctor} \Mapsto \emptyset][x_{pred} \Mapsto \emptyset][x_{acc} \Mapsto \emptyset]\,|\,\square}$$

**Figure 17** Typing Judgement

**Type Checking Extracted Methods**

T-APP

$$\Gamma \vdash e_1 : (x{:}\tau \to \exists X.R\,;\, \psi_{1+} \,|\, \psi_{1-}\,;\, \emptyset) \,|\, \overrightarrow{T_1}$$
$$\Gamma, \psi_{1+} \vdash e_2 : (\sigma\,;\, \psi_{2+} \,|\, \psi_{2-}\,;\, o_2) \,|\, \overrightarrow{T_2}$$
$$\underline{\Gamma \vdash \sigma <: \tau \quad X \,\#\, \sigma \quad X \,\#\, \psi_{1+} \quad X \,\#\, \Gamma}$$
$$\Gamma \vdash (e_1\, e_2) : R[x \stackrel{\sigma}{\Longmapsto} o_2] \,|\, \overrightarrow{T_1} + \overrightarrow{T_2}$$

T-CONS

$$\Gamma \vdash e_1 : (\tau_1\,;\, \mathbb{TT} \,|\, \mathbb{TT}\,;\, o_1) \,|\, \overrightarrow{T_1}$$
$$\underline{\Gamma, \overrightarrow{T_1}, \vdash e_2 : (\tau_2\,;\, \mathbb{TT} \,|\, \mathbb{TT}\,;\, o_2) \,|\, \overrightarrow{T_2}}$$
$$\Gamma \vdash (\mathbf{cons}\, e_1\, e_2) : (\tau_1 \times \tau_2\,;\, \mathbb{TT} \,|\, \mathbb{TT}\,;\, \emptyset) \,|\, \overrightarrow{T_1} + \overrightarrow{T_2}$$

T-FST

$$\Gamma \vdash e : (\tau_1 \times \tau_2\,;\, \mathbb{TT} \,|\, \mathbb{TT}\,;\, o) \,|\, \overrightarrow{T}$$
$$\underline{R = (\tau_1\,;\, \mathbb{TT} \,|\, \mathbb{TT}\,;\, (\mathbf{fst}\, x))}$$
$$\Gamma \vdash (\mathbf{fst}\, e) : R[x \stackrel{\tau_1}{\Longmapsto} o] \,|\, \overrightarrow{T}$$

T-SND

$$\Gamma \vdash e : (\tau_1 \times \tau_2\,;\, \mathbb{TT} \,|\, \mathbb{TT}\,;\, o) \,|\, \overrightarrow{T}$$
$$\underline{R = (\tau_2\,;\, \mathbb{TT} \,|\, \mathbb{TT}\,;\, (\mathbf{snd}\, x))}$$
$$\Gamma \vdash (\mathbf{snd}\, e) : R[x \stackrel{\tau_2}{\Longmapsto} o] \,|\, \overrightarrow{T}$$

T-IEXI

$$X \,\#\, \Gamma \quad X \,\#\, \overrightarrow{T}$$
$$\underline{\Gamma \vdash v : R[X \Longmapsto \tau] \,|\, \overrightarrow{T}}$$
$$\Gamma \vdash v : R \,|\, \overrightarrow{T}, X$$

**Figure 18** Typing Judgement Continued

* **Subcase T-NAT**: $\Gamma \vdash n : (\mathbf{N}\,;\, \mathbb{TT} \,|\, \mathbb{FF}\,;\, \emptyset) \,|\, \square$
    1. $o = \emptyset$
    2. since $n \neq \mathit{false}$, by M-TOP $\rho \models \Gamma, \mathbb{TT}$ trivially.
    3. By assumption, $\vdash n : (\mathbf{N}\,;\, \mathbb{TT} \,|\, \mathbb{FF}\,;\, \emptyset) \,|\, \square$
* **Subcase T-IEXI**:
  By inversion, $\Gamma \vdash n : (\mathbf{N}\,;\, \mathbb{TT} \,|\, \mathbb{FF}\,;\, \emptyset) \,|\, \square, X$
  The rest follows the same argument to the previous subcase.
- **Case v = $sn(v : \tau, \overrightarrow{sp\,v})$**:
  Two rules can be derived as the last one in the typing derivation:
  T-STRUCT-INSTANCE, T-IEXI. Proceed by cases.
    * **Subcase T-STRUCT-INSTANCE**:
        1. $o = \emptyset$
        2. since $sn(v : \tau, \overrightarrow{sp\,v}) \neq \mathit{false}$, by M-TOP $\rho \models \Gamma, \mathbb{TT}$ trivially.
        3. By assumption, $\Gamma \vdash sn(v : \tau, \overrightarrow{sp\,v_p}) : (sn(\tau, \overrightarrow{sp})\,;\, \mathbb{TT} \,|\, \mathbb{FF}\,;\, \emptyset) \,|\, \square$
    * **Subcase T-IEXI**: follow a similiar argument to subcase T-IEXI in **Case v = n**.
- The rest cases follow an similar argument.
- **B-Var** $\rho \vdash x \Downarrow v$
  Two rules can be derived as the last one in valid typing derivation: T-VAR.
  By inversion, $\Gamma \vdash x \in \tau$

1. By inversion on the evaluation derivation, $v = \rho(x)$
2. To show $\rho \models x \in \mathsf{F}$ or $\rho \models x \notin \mathsf{F}$:
   By M-Type, if $v = \mathsf{false}$, $\rho \models x \in \mathsf{F}$. Otherwise, $v \neq false$. Do structural induction on $v$.
   a. subcase: $v = n$. By M-NOT-TYPE, since $\vdash v : (\mathbf{N} ; \psi'_+ \mid \psi'_- ; o') \mid \overrightarrow{T}$ and there is no overlap between an $\mathbf{N}$ and an $\mathsf{F}$, $\rho \models \Gamma, x \notin \mathsf{F}$
   b. the rest subcases follow a similar argument
3. Since $\Gamma \vdash x \in \tau$, by Lemma 1, $\rho \models x \in \tau$. Then by inversion on M-Type, $\vdash v : (\tau ; \psi'_+ \mid \psi'_- ; o') \mid \overrightarrow{T}$ for some $\psi'_+, \psi'_-$ and $o'$

- **B-Abs** $\rho \vdash \lambda x{:}\tau.e \Downarrow [\rho, \lambda x{:}\tau.e]$
  $\Gamma \vdash \lambda x{:}\tau.e : (\exists X. x{:}\tau \to R ; \mathbb{TT} \mid \mathbb{FF} ; \emptyset) \mid \overrightarrow{T}, R = (\tau_o ; \psi_{f+} \mid \psi_{f-} ; o)$
  By inversion on the typing rule: $\Gamma, X, x \in \tau \vdash e : R \mid \overrightarrow{T}$.

  1. $o = \emptyset$
  2. By lemma 1 and because $[\rho, \lambda x{:}\tau.e] \notin \mathsf{F}$, $\rho \models \Gamma, \mathbb{TT}$
  3. By assumption and T-Closure, $\Gamma \vdash [\rho, \lambda x{:}\tau.e] : (\exists X. x{:}\tau \to R ; \psi_+ \mid \mathbb{FF} ; \emptyset) \mid \overrightarrow{T}$

- **B-Struct-Related-Operation** $\rho \vdash e_1 \Downarrow so, \rho \vdash e_2 \Downarrow v_1, \delta_s(so, v_1) = v$
  The valid typing derivation is T-App: $e = (e_1 \, e_2)$, $o = o_f[x \overset{\sigma}{\mapsto} o_2]$, $\psi_+ = \psi_{f+}[x \overset{\sigma}{\mapsto} o_2]$, $\psi_- = \psi_{f-}[x \overset{\sigma}{\mapsto} o_2]$
  By inversion on T-App, we know:
  - $\Gamma \vdash e_1 : (\exists X. x{:}\sigma \to R ; \psi_{1+} \mid \psi_{1-} ; o_1) \mid \overrightarrow{T_1}$
  - $\Gamma, \overrightarrow{T_1}, \psi_{1+} \vdash e_2 : (\sigma_2 ; \psi_{2+} \mid \psi_{2-} ; o_2) \mid \overrightarrow{T_2}$
  - $\Gamma \vdash \sigma_2 <: \sigma$
  - $R = (\tau_f ; \psi_{f+} \mid \psi_{f-} ; o_f)$
  - $\overrightarrow{T} = \overrightarrow{T_1} + \overrightarrow{T_2}$

  Doing induction on $so$. Proceed by cases:

  1. $so = \mathsf{ctor}(sn, \sigma, \overrightarrow{sp\,v_p})$ and $v = sn(v_1 : \tau, \overrightarrow{sp\,v_p})$:
     By applying IH to $\Gamma \vdash e_1 : (x{:}\sigma \to \exists X.R ; \psi_{1+} \mid \psi_{1-} ; o_1) \mid \overrightarrow{T_1}$ and $\rho \vdash e_1 \Downarrow so$, we know:
     - $\Gamma \vdash \mathsf{ctor}(sn, \tau, \overrightarrow{sp\,v_p}) : (x{:}\sigma \to \exists X.R ; \mathbb{TT} \mid \mathbb{FF} ; \emptyset) \mid \overrightarrow{T_1}$
     - $R = (sn(\sigma, \overrightarrow{sp}) ; \mathbb{TT} \mid \mathbb{FF} ; \emptyset)$
     - X doesn't appear anywhere in the bodies.

     Then we are able to show:
     a. $o = \emptyset$
     b. since $sn(v_1 : \tau, \overrightarrow{sp\,v_p}) \neq \mathsf{false}$, $\rho \models \mathbb{TT}$ trivially by M-TOP
     c. $\vdash sn(v_1 : \tau, \overrightarrow{sp\,v_p}) : (sn(\sigma, \overrightarrow{sp}) ; \mathbb{TT} \mid \mathbb{FF} ; \emptyset) \mid \overrightarrow{T}$
  2. $so = \mathsf{acc}(sn(\sigma, \overrightarrow{sp}) v_p)$ and $v_1 = sn(v : \sigma, \overrightarrow{sp\,v_p})$:
     By applying IH to $\Gamma \vdash e_1 : (x{:}\sigma \to \exists X.R ; \psi_{1+} \mid \psi_{1-} ; o_1) \mid \overrightarrow{T_1}$ and $\rho \vdash e_1 \Downarrow so$, we know:
     - $\Gamma \vdash \mathsf{acc}(sn(\tau, \overrightarrow{sp\,v_p})) : (x{:}sn(\tau, \overrightarrow{sp\,v_p}) \to \exists X.R ; \mathbb{TT} \mid \mathbb{FF} ; \emptyset) \mid \overrightarrow{T}$
     - $R = (\sigma ; \mathbb{TT} \mid \mathbb{TT} ; \emptyset)$

## Type Checking Extracted Methods

- X doesn't appear anywhere in the bodies.
a. $o = \emptyset$
b. either $v =$ false or $v \neq$ false, by M-TOP, $\rho \models \Gamma, \mathbb{TT}$.
c. $\vdash v : (\tau \,;\, \mathbb{TT} \,|\, \mathbb{TT} \,;\, \emptyset) \,|\, \overrightarrow{T}$

3. $so = \text{p-acc}(sp_i, \tau_{p_i})$, $v_1 = sn(v_a : \sigma, \overrightarrow{sp\,v_p})$, and $v = v_{p_i}$
   By applying IH to $\Gamma \vdash e_1 : (x{:}\sigma \to \exists X.R \,;\, \psi_{1+} \,|\, \psi_{1-} \,;\, o_1) \,|\, \overrightarrow{T_1}$ and $\rho \vdash e_1 \Downarrow so$,
   we know:
   - $\Gamma \vdash \text{p-acc}(sp_i, \tau_{p_i}) : (x{:}\textbf{Has-Prop}(sp_i) \to \exists Self.R \,;\, \psi'_+ \,|\, \psi_- \,;\, o') \,|\, \overrightarrow{T_1}$
   - $R = (\tau_{p_i} \,;\, x \in Self \,|\, \mathbb{TT} \,;\, o_f)$
   - $\Gamma \vdash v_{p_i} : R[o_2 \mapsto x][Self \mapsto sn(\tau_s, \overrightarrow{sp\,\tau_p})] \,|\, \overrightarrow{T_v}$
   a. if $o_2 = \emptyset$ or $o_f = \emptyset$, $o = \emptyset$. Otherwise, $o \neq \emptyset$. Since $\rho(o_f) = v_{p_i}$, x is absent and the variable in $o_2$ is also bound in $\rho$, $\rho(o) = v_{p_i}$.
   b. **Subcase** $v \neq$ false : Since the variable in $o_2$ is bound in $\rho$, $\rho \models \Gamma, o_2 \in Self$.
      **Subcase** $v =$ false : $\rho \models \Gamma, \mathbb{TT}$ by M-Top.
   c. by IEXI, $\Gamma \vdash v_{p_i} : R[o_2 \mapsto x] \,|\, \overrightarrow{T_v}, Self$

4. The rest subcases follow a similar argument to the previous subcases

- **B-Beta** $\rho \vdash e_1 \Downarrow [\rho_c, \lambda x{:}\tau_c.e_c]$, $\rho \vdash e_2 \Downarrow v_2$, $\rho_c[x := v_2] \vdash e \Downarrow v$,
  The last rule in the typing derivation is T-App: $e = (e_1\,e_2)$, $o = o_f[x \xRightarrow{\sigma} o_2]$, $\psi_+ = \psi_{f+}[x \xRightarrow{\sigma} o_2]$, $\psi_- = \psi_{f-}[x \xRightarrow{\sigma} o_2]$
  By inversion on T-App, we know:
  - $\Gamma \vdash e_1 : (x{:}\sigma \to \exists X.R \,;\, \psi_{1+} \,|\, \psi_{1-} \,;\, o_1) \,|\, \overrightarrow{T_1}$
  - $\Gamma, \overrightarrow{T_1}, \psi_{1+} \vdash e_2 : (\sigma_2 \,;\, \psi_{2+} \,|\, \psi_{2-} \,;\, o_2) \,|\, \overrightarrow{T_2}$
  - $\Gamma \vdash \sigma_2 <: \sigma$
  - $R = (\tau_f \,;\, \psi_{f+} \,|\, \psi_{f-} \,;\, o_f)$
  - $\overrightarrow{T} = \overrightarrow{T_1} + \overrightarrow{T_2}$

  By IH on $\Gamma \vdash e_1 : (x{:}\sigma \to \exists X.R \,;\, \psi_{1+} \,|\, \psi_{1-} \,;\, o_1) \,|\, \overrightarrow{T_1}$ and $\rho \vdash e_1 \Downarrow [\rho_c, \lambda x{:}\tau_c.e_c]$,
  $\Gamma \vdash [\rho_c, \lambda x{:}\sigma.e_c] : (x{:}\sigma \to \exists X.R \,;\, \psi_{1+} \,|\, \psi_{1-} \,;\, o_1) \,|\, \overrightarrow{T_1}$,
  By inversion on T-Closure,
  $\exists \Gamma'. \rho_c \models \Gamma'$ and $\Gamma' \vdash \lambda x{:}\sigma.e : (x{:}\sigma \to \exists X.R \,;\, \psi_{1+} \,|\, \psi_{1-} \,;\, o_1) \,|\, \overrightarrow{T_1}$
  By inversion on T-Abs, $\Gamma', x \in \sigma \vdash e_f : (\tau \,;\, \psi_{f+} \,|\, \psi_{f-} \,;\, o_f) \,|\, \overrightarrow{T_1}$

  1. if $o_2 = \emptyset$ or $o_f = \emptyset$, $o = \emptyset$. Otherwise, $o_2 \neq \emptyset$. Extend $\rho_c$ with $o_2$ and $v_2$, and substitute x in $o_f$ : $\rho_c[o_2 := v_2](o) = v$. Since x is no longer present in $\rho_c$ and $o$ and the free variable in $o$ is also bound in $\rho$, $\rho(o) = v$.
  2. By applying IH to $\rho_c[x := v_2] \vdash e \Downarrow v$ and $\Gamma', x \in \sigma \vdash e_f : (\tau \,;\, \psi_{f+} \,|\, \psi_{f-} \,;\, o_f) \,|\, \overrightarrow{T_1}$, if $v \neq$ false then $\rho_c[x := v_2] \models \Gamma', \psi_{f+}$, or $v =$ false then $\rho_c[x := v_2] \models \Gamma', \psi_{f-}$.
     If $o_2 = \emptyset$, by subsitutition, $\psi_{f+}[o_2 \mapsto x] = \mathbb{TT}$ if $v \neq$ false or $\psi_{f-}[o_2 \mapsto x] = \mathbb{TT}$. Then $\rho \models \Gamma, \mathbb{TT}$ by M-TOP trivially.
     Otherwise, $o_2 \neq \emptyset$. In this case, by substitution, x doesn't appear in $\psi_{f+}[o_2 \mapsto x]$ or $\psi_{f-}[o_2 \mapsto x]$ any more. Extend $\rho_c$ with $o_2$ and $v_2$: $\rho_c[o_2 := v_2] \models \Gamma', \psi_{f+}[o_2 \mapsto x]$ or $\rho_c[o_2 := v_2] \models \Gamma', \psi_{f-}[o_2 \mapsto x]$. Since the variable in $o_2$ is bound $\rho$ such that $\rho(o_2) = v_2$ and also it is well typed under $\Gamma$, $\rho \models \Gamma, \psi_{f+}[o_2 \mapsto x]$ or

6:40

$\rho \models \Gamma, \psi_{f-}[o_2 \Mapsto x]$. Since $\Gamma, x \in \sigma \vdash \psi_{f+}$ or $\Gamma, x \in \sigma \vdash \psi_{f-}$ and after substitution x doesn't exist any more, $\Gamma \vdash \Gamma, \psi_{f+}[o_2 \Mapsto x]$ or $\Gamma \vdash \Gamma, \psi_{f-}[o_2 \Mapsto x]$ by Lemma 3. By Lemma 1, $\rho \models \Gamma, \psi_{f+}[o_2 \Mapsto x]$ or $\rho \models \Gamma, \psi_{f-}[o_2 \Mapsto x]$.

3. By IH, $\vdash v : (\tau ; \psi_+ | \psi_- ; o) | \overrightarrow{T}$

- **B-IF-TRUE** $\rho \vdash e_1 \Downarrow v_1, v_1 \neq \text{false}, \rho \vdash e_2 \Downarrow v$
  The last rule in the valid typing derivation is T-IF: $e = (\mathbf{if}\, e_1\, e_2\, e_3)$, $\psi_+ = \psi_{2+} \vee \psi_{3+}$, $\psi_- = \psi_{2-} \vee \psi_{3-}$
  By inversion, we know
  - $\Gamma \vdash e_1 : (\top ; \psi_{1+} | \psi_{1-} ; o') | \overrightarrow{T_1}$
  - $\Gamma, \overrightarrow{T_1}, \psi_{1+} \vdash e_2 : (\tau ; \psi_{2+} | \psi_{2-} ; o) | \overrightarrow{T_2}$
  - $\Gamma, \overrightarrow{T_1}, \psi_{1-} \vdash e_3 : (\tau ; \psi_{3+} | \psi_{3-} ; o) | \overrightarrow{T_3}$
  - $\overrightarrow{T} = \overrightarrow{T_1} + \overrightarrow{T_2} + \overrightarrow{T_3}$

  By IH on $\rho \vdash e_1 \Downarrow v_1$ and $\Gamma \vdash e_1 : (\top ; \psi_{1+} | \psi_{1-} ; \emptyset) | \overrightarrow{T_1}$, $\rho \models \Gamma, \psi_{1+}$
  By IH on $\rho \vdash e_2 \Downarrow v$ and $\Gamma, \overrightarrow{T_1}, \psi_{1+} \vdash e_2 : (\tau ; \psi_{2+} | \psi_{2-} ; o) | \overrightarrow{T_2}$, we are able to prove the following:
  1. $o = \emptyset$ or $\rho(o) = v$
  2. if $v \neq false$, since $\rho \models \Gamma, \psi_{1+}$, by Lemma 1 $\rho \models \psi_{2+}$. By M-OR, $\rho \models \psi_{2+} \vee \psi_{3+}$.
  3. By IH, $\vdash v : (\tau ; \psi_+ | \psi_- ; o) | \overrightarrow{T}$

- **B-IF-False** $\rho \vdash e_1 \Downarrow v_1, v_1 = \text{false}, \rho \vdash e_3 \Downarrow v$ Follow an argument similar to **B-IF-TRUE** while doing IH on the else branch.

- **B-Let** $\rho \vdash e_1 \Downarrow v_1, \rho[x := v_1] \vdash e_2 \Downarrow v$
  The last rule in the typing derivation is T-LET: $e = (\mathbf{let}\,(x\, e_1)\, e_2)$, $o = o_2[x \Mapsto o_1]$, $\psi = \psi_{2+}[x \Mapsto o_1]$, $\psi = \psi_{2-}[x \Mapsto o_1]$
  By inversion on this rule, we know
  - $\Gamma \vdash e_1 : (\tau_1 ; \psi_{1+} | \psi_{1-} ; o_1) | \overrightarrow{T_1}$
  - $\psi_x = (x \notin \mathsf{F} \wedge \psi_{1+}) \vee (x \in \mathsf{F} \wedge \psi_{1-})$
  - $\Gamma, \overrightarrow{T_1}, x \in \tau_1, x \equiv o_1, \psi_x \vdash e_2 : R | \overrightarrow{T_2}$
  - $R = (\tau_2 ; \psi_{2+} | \psi_{2-} ; o_2)$

  By applying IH to $\rho \vdash e_1 \Downarrow v_1$ and $\Gamma \vdash e_1 : (\tau_1 ; \psi_{1+} | \psi_{1-} ; o_1) | \overrightarrow{T_1}$, $\rho \models \psi_{1+}$ or $\rho \models \psi_{1-}$.
  By applying IH to $\rho[x := v_1] \vdash e_2 \Downarrow v$ and $\Gamma, \overrightarrow{T_1}, x \in \tau_1, x \equiv o_1, \psi_x \vdash e_2 : R | \overrightarrow{T_2}$, $\rho[x := v_1] \models \psi_{2+}$ or $\rho[x := v_1] \models \psi_{2-}$. Then we can show:
  1. if $o_1 = \emptyset$ or $o_2 = \emptyset$, $o = \emptyset$. Otherwise, $o_1 \neq \emptyset$. Since $\rho[x := v_1](o_2) = v$, by substituting x with $o_1$ in $o_2$, $\rho[x := v_1](o_2[x \Mapsto o_1]) = v$. Since $x \equiv o_1$, $\rho(o_1) = v_1$ by M-ALIAS. Because the variable in $o_1$ is already bound in $\rho$, $\rho(o_2[x \Mapsto o_1]) = v$
  2. if $v = $ false, we need to show: $\rho \models \Gamma, \psi_{2+}[x \Mapsto o_1]$. Since $\rho[x := v_1] \models \Gamma, \psi_{2+}$, $\rho[x := v_1] \models \Gamma, \psi_{2+}[x \Mapsto o_1]$ by substituting x with $o_1$. Since $x \equiv o_1$, $\rho(o_1) = v_1$ by M-ALIAS. Because the variable in $o_1$ is already bound in $\rho$, $\rho \models \Gamma, \psi_{2+}[x \Mapsto o_1]$
  3. By IH, $\vdash v : (\tau ; \psi_+ | \psi_- ; o) | \square$

- **B-Let-Struct-Property** $v_p = \mathsf{pd}(sp), v_{pred} = \mathsf{p\text{-}pred}(sp), v_{acc} = \mathsf{p\text{-}acc}(sp, \tau), \rho[sp := v_p][x_{pred} := v_{pred}][x_{acc} := v_{acc}] \vdash e \Downarrow v$
  This case can be trivially proved by IH

**Type Checking Extracted Methods**

- **B-Let-Struct** $v_p = \mathsf{pd}(sp), v_{pred} = \mathsf{p\text{-}pred}(sp), v_{acc} = \mathsf{p\text{-}acc}(sp, \tau), \rho[sp := v_p][x_{pred} := v_{pred}][x_{acc} := v_{acc}] \vdash e \Downarrow v$
  This case can be trivially proved by IH
- **B-Fst** or **B-Snd**: $\rho \vdash e \Downarrow \langle v_1, v_2 \rangle$
  These two cases can be trivially proved by IH
- **B-Pair**: $\rho \vdash e_1 \Downarrow v_1, \rho \vdash e_2 \Downarrow v_2$
  Trivially proved by IH and subsumption.
- **B-Prim**: $\rho \vdash e_1 \Downarrow op, \rho \vdash e_2 \Downarrow v_2, \delta(op, v_2) = v$
  The valid typing derivation is T-App: $\Gamma \vdash (e_1 \, e_2) : (\tau \,;\, \psi_+ \,|\, \psi_- \,;\, o_f) \,|\, \overrightarrow{T}$.
  Do case-by-case proof on $op$ and follow an argument similar to **B-Struct-Values**

□

**Lemma 3.** *If $\Gamma, \psi' \vdash \psi$, then $\Gamma, \psi' \vdash \Gamma, \psi$*

*Proof.* By definition, $\Gamma, \psi' \vdash \Gamma$. By M-AndI, $\Gamma, \psi' \vdash \Gamma \wedge \psi$. Since $\Gamma \wedge \psi = \Gamma, \psi$, $\Gamma, \psi' \vdash \Gamma, \psi$

□

**About the authors**

**Yuquan Fu** is a Ph.D. student at Indiana University. Contact him at yuqfu@iu.edu.

**Sam Tobin-Hochstadt** is an associate professor at Indiana University. Contact him at samth@cs.indiana.edu.